\documentclass{nature}
\usepackage{amsmath}
\usepackage[T1]{fontenc}
\usepackage[latin1]{inputenc}
\usepackage{graphicx}
\usepackage{amssymb}
\usepackage{dcolumn}
\usepackage{color}
\usepackage{bm}
\usepackage[normalem]{ulem}
\usepackage{pdfpages}
\usepackage{pgffor}
\usepackage{url}
\usepackage{babel}

\begin{document}

\title{Observation of ballistic upstream modes at 
fractional quantum Hall edges of graphene}
\author{Ravi Kumar$^{1}$\footnote{equally contributed}, Saurabh Kumar Srivastav$^{1}$\footnote{equally contributed}, Christian Sp{\r a}nsl{\"a}tt$^{2,3,4}$, K. Watanabe$^{5}$, T. Taniguchi$^{5}$, Yuval Gefen$^{6}$, Alexander D. Mirlin$^{3,4,7,8}$, and Anindya Das$^{1}$\footnote{anindya@iisc.ac.in}}
\maketitle

\begin{affiliations}
\item Department of Physics, Indian Institute of Science, Bangalore, 560012, India.
\item Department of Microtechnology and Nanoscience (MC2),Chalmers University of Technology, S-412 96 G\"oteborg, Sweden.
\item Institute for Quantum Materials and Technologies, Karlsruhe Institute of Technology, 76021 Karlsruhe, Germany.
\item Institut f{\"u}r Theorie der Kondensierten Materie, Karlsruhe Institute of Technology, 76128 Karlsruhe, Germany.
\item National Institute of Material Science, 1-1 Namiki, Tsukuba 305-0044, Japan.
\item Department of Condensed Matter Physics, Weizmann Institute of Science, Rehovot 76100, Israel.
\item Petersburg Nuclear Physics Institute, 188300 St. Petersburg, Russia.
\item L. D. Landau Institute for Theoretical Physics RAS, 119334 Moscow, Russia.
\end{affiliations}

\noindent\textbf{The structure of edge modes at the boundary of quantum Hall (QH) phases forms the basis for understanding low energy transport properties. 
In particular, the presence of ``upstream'' modes, moving against the direction of charge current flow, is critical for the emergence of renormalized modes with exotic quantum statistics. Detection of excess noise at the edge is a smoking gun for the presence of upstream modes. Here we report on noise measurements at the edges of fractional QH (FQH) phases realized in dual graphite-gated bilayer graphene devices. 
A noiseless dc current is injected at one of the edge contacts, and the noise generated at contacts at $L= 4\,\mu$m or $10\,\mu$m away along the upstream direction is studied. For integer and particle-like FQH states, no detectable noise is measured. By contrast, for ``hole-conjugate''  FQH states, we detect a strong noise proportional to the injected current, unambiguously proving the existence of upstream modes. The noise magnitude remaining independent of length together with a remarkable agreement with our theoretical analysis demonstrates the ballistic nature of upstream energy transport, quite distinct from the diffusive propagation reported earlier in GaAs-based systems. Our investigation opens the door to the study of upstream transport in more complex geometries and in edges of non-Abelian phases in graphene.}

\begin{figure*}
\centerline{\includegraphics[width=1.0\textwidth]{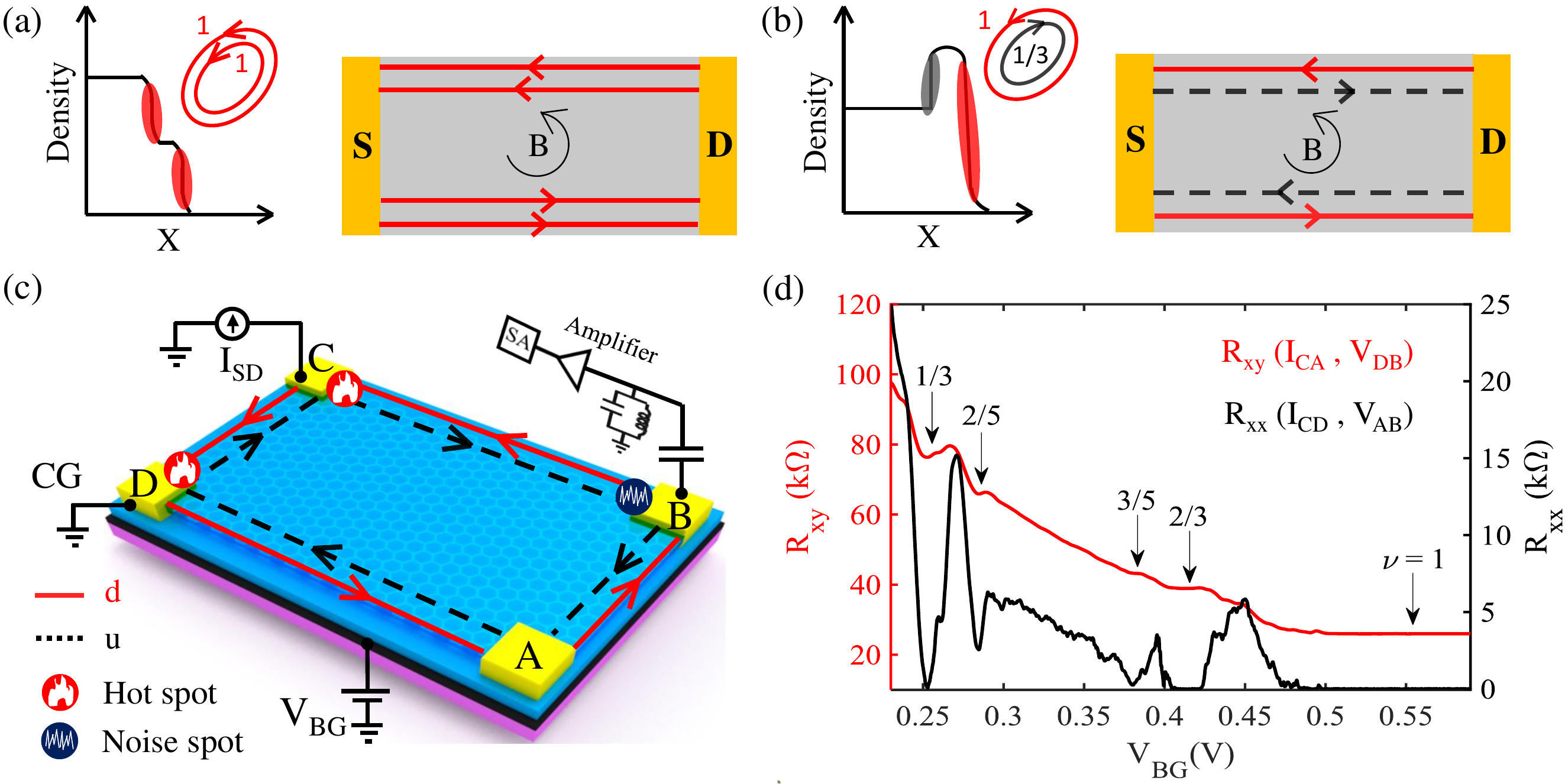}}
\caption{\textbf{Edge profile, Device schematic, Hot spot, Noise spot, and QH response.} (\textbf{a}) and (\textbf{b}) show the density profile at the edge of a sample with co-propagating and counter-propagating edge structures for $\nu=2$ and for hole-conjugate $\nu=2/3$ states, respectively. x is a coordinate across the edge. 
Downstream modes are red, while upstream modes are black. 
(\textbf{c}) Schematic of device and measurement setup, where the device is set into $\nu = 2/3$ FQH state. A noiseless dc current is injected at contact C and terminates into cold ground (CG) contact D. 
The upstream mode carry the heat from the hot spot near the contact C and create the noise spot at the contact B, 
which is measured at a frequency of $\sim$770\,kHz using a LCR resonance circuit followed by amplifier chain and spectrum analyzer. (\textbf{d}) Transverse Hall resistance ($R_{xy}$) and longitudinal resistance ($R_{xx}$) of the device-1 at $10$\,T. 
}
\label{Figure1}
\end{figure*}

\textbf{Introduction:} Transport in integer quantum Hall (QH) states occurs through one-dimensional edge modes located at the edge of the sample with downstream chirality dictated by the magnetic field (Fig.~\ref{Figure1}a). This is also true for particle-like fractional quantum Hall (FQH) states~\cite{beenakker1990edge,wen1990chiral}. By contrast, so called ``hole-conjugate'' FQH states ($i+1/2< \nu <i+1$ with filling factor $\nu$ and $i$=0,1,2,..) are expected to host counter-propagating chiral edge modes moving respectively along the downstream  and upstream directions. A paradigmatic example is the $\nu=2/3$ bulk state. MacDonald and Johnson~\cite{PhysRevLett.64.220,johnson1991composite} proposed that the edge supports two counter-propagating modes: a downstream mode, $\nu=1$, and an upstream $\nu=1/3$ mode (Fig.~\ref{Figure1}b). The existence of upstream modes is of fundamental importance and crucially affects transport properties including electrical and thermal transport~\cite{altimiras2012chargeless,jezouin2013quantum}, noise, and particle interferometry~\cite{ji2003electronic,nakamura2019aharonov}.
 
There has been an 
extensive effort over recent years to detect experimentally upstream modes and their properties in GaAs/AlGaAs quantum well based 2DEG. 
Two questions then come to mind. The first is whether the upstream modes can be detected by measuring the electrical conductance. Indeed, for the hole-conjugate $\nu=2/3$ state, and for distances much smaller than the charge equilibration length, the two-terminal electric conductance $G$ is expected~\cite{protopopov2017transport} to be $4/3\frac{e^2}{h}$ (instead of  $2/3\frac{e^2}{h}$), confirming the counter propagating character of edge modes. This value was indeed measured, but only on engineered edges at interfaces between two FQH states~\cite{Cohen2019}. In experiments on a conventional edge (the boundary of a $\nu=2/3$ FQH state), $G = 2/3\frac{e^2}{h}$ is found to be robust and essentially equal to the filling factor $\nu = 2/3$, implying that charge propagates only downstream direction; 
one cannot tell anything about the presence or absence of upstream modes~\cite{protopopov2017transport,nosiglia2018incoherent,spaanslatt2021contacts}. The second question is whether the upstream mode can be detected by measuring the thermal conductance. Thermal transport on an edge may be qualitatively different from  charge transport \cite{protopopov2017transport,nosiglia2018incoherent}. Even if charge propagates only downstream, energy may propagate upstream. Measurements of thermal conductance $G_Q$ at $\nu=2/3$ and related fillings yielded results fully consistent with the theory expected from hole-conjugated character of these states with upstream modes~\cite{banerjee2017observed,banerjee2018observation,srivastav2021vanishing,melcer2021absent}. 
Still, measurements of $G_Q$ do not prove unambiguously the  existence of upstream modes. Ideally, for $\nu=2/3$, hole-conjugate state with counter-propagating 1 and 1/3 modes is favored by particle-hole symmetry, which is however only approximate in any realistic situation. Note, however, that an edge with two co-propagating $\nu=1/3$ modes is
a fully legitimate candidate~\cite{wen1990chiral,beenakker1990edge} too. Such a model would be consistent not only with measured electrical conductance $G = 2/3\frac{e^2}{h}$ but also with the thermal conductance measurements reported in Ref.~\cite{srivastav2021vanishing}. On a broader scope, several alternative approaches attempting to establish the presence of upstream current of heat at edges of a variety of FQH states in GaAs/AlGaAs structures were employed in Refs.~\cite{bid2010observation,Dolev2011characterizing,Gross2012upstream,Venkatachalam2012local,Gurman2012extracting}. Those studies used structures involving quantum point contacts or quantum dots. It is also worth noting that candidates to the non-Abelian $\nu=5/2$ state possess different number of upstream modes, and intensive current efforts aim at understanding which of them is actually realized in experiment~\cite{banerjee2018observation,ma2019partial,asasi2020partial,simon2020partial,park2020noise,dutta2021novel,yutushui2021identifying}. 

The emergence of the two-dimensional graphene platform opened up a new era in the study of FQH physics  \cite{li2017even,kim2019even}, with different Landau level structure (as compared with traditional GaAs heterostructures), new fractions, and enriched family of quantum-Hall states due to an interplay of spin, orbital, and valley degrees of freedom~\cite{yang2006collective,sodemann2014broken,kharitonov2012phase,weitz2010broken,young2012spin,kumar2018equilibration}. Furthermore, graphene features an unprecedented sharp confining potential and is thus expected to exhibit bulk-edge correspondence without additional complex edge reconstruction~\cite{li2013evolution,hu2011realizing}. Also, in view of the sharp edge, one expects strong interaction between the edge modes in graphene, which may give access to regimes that are difficult to reach in GaAs structures. Remarkably, for graphene or graphene-based hybrid structures, no direct evidence for the presence of upstream modes has so far been reported. Here, we, for the first time, report a smoking gun signature of upstream modes for hole-conjugate FQH states in graphene and identify their nature, employing noise spectroscopy, which is a purely electrical tool. The essence of our approach is as follows \cite{spaanslatt2019topological}. When a bias is applied to a FQH edge segment, the Joule heat is 
dissipated at the ``hot spots'' as shown in Fig. 1c. In the presence of upstream modes, heat is transported upstream to the so-called noise spot (Fig.~\ref{Figure1}c), where the heat partitions the charge current and thereby generates noise. 

We have carried out electrical conductance together with noise measurements at integer QH states, electron-like state $\nu =1/3$ and hole-conjugate states $\nu =2/3$ and $\nu =3/5$, realized in a dual graphite-gated hexagonal-boron-nitride (hBN)-encapsulated high-mobility bilayer graphene (BLG) devices in a cryo-free dilution fridge at base temperature of $\sim 20$ mK (electron temperature $\sim 28$ mK). For $\nu =1/3$ state and integer QH states, we do not detect any excess noise along the upstream direction. This is expected, because the corresponding edge states do not host upstream modes. By contrast, for $\nu=2/3$ and $\nu =3/5$ FQH states, a finite noise is detected which increases with increasing injected current. At the same time, the averaged current in upstream direction is zero. Thus, noise detection unambiguously demonstrates that upstream modes exist for the hole-conjugate FQH states in graphene and only carry heat energy. 
Moreover, the magnitude of the noise remains constant for two different lengths $L= 10\,\mu$m and $4\,\mu$m between the current injecting contact and noise detection point in the same device. 
Moreover, our experimentally measured noise magnitude matches remarkably well with our theoretical analysis. 
This conclusively demonstrates the ballistic nature of upstream modes, implying the absence of thermal equilibration on the length scales employed in the experiment. 

\begin{figure}[!htb]
\centerline{\includegraphics[width=1.0\textwidth]{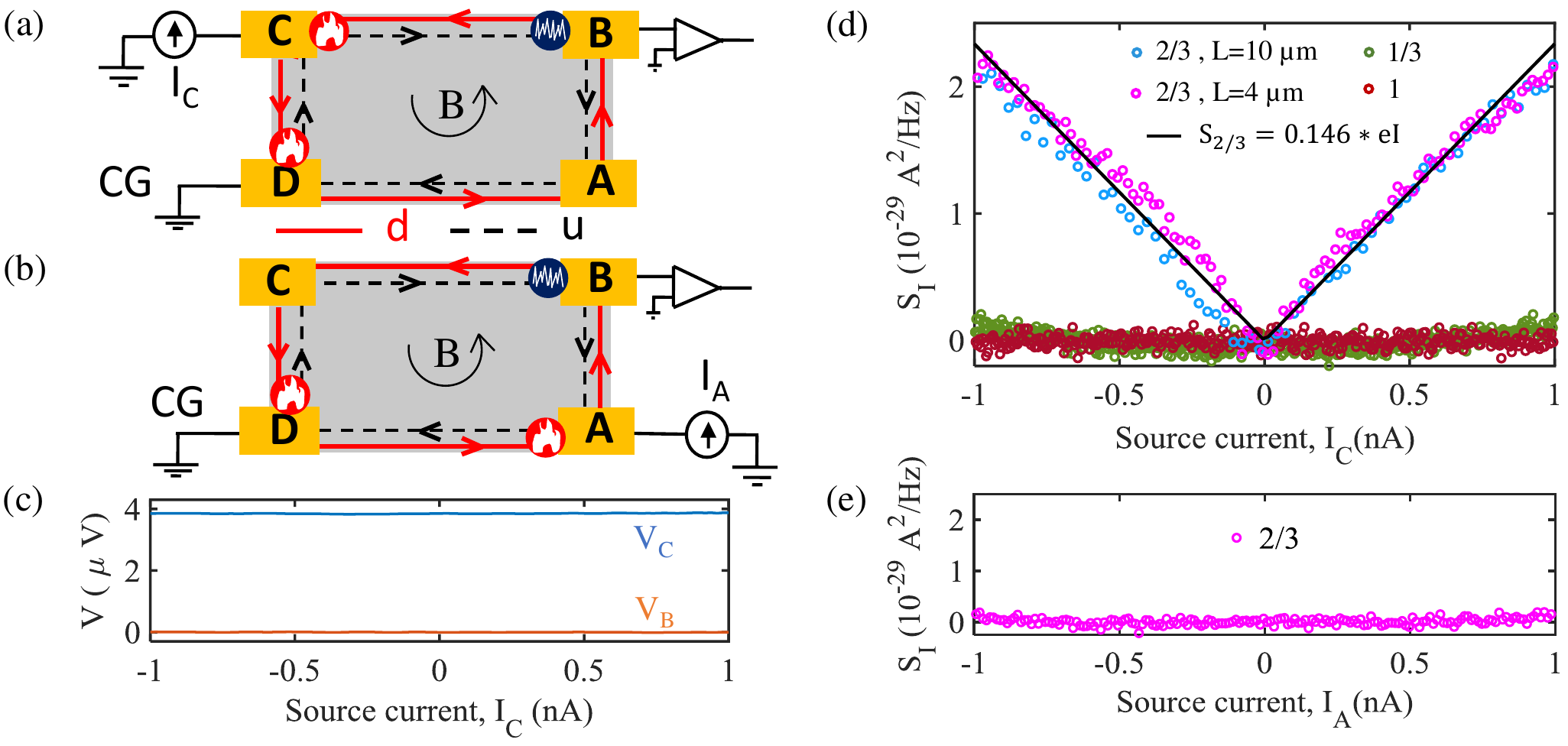}}
\caption{\textbf{Noise data} (\textbf{a}) and (\textbf{b}) show the noise detection scheme 
of the device-1 while the current is injected at contact C and contact A, respectively. 
(\textbf{c}) Differential bias response of $\nu=2/3$ state, where a 100\, pA AC signal on top of the DC bias current is injected at contact C and the voltages are measured at contact C ($V_C$) and B ($V_B$) using Lock-In amplifier. 
(\textbf{d}) Noise measured along the upstream direction for longer length ($10\,\mu$m) as shown in Fig. 2a for $\nu=1$ (brown circles), $\nu=1/3$ (olive circles) and $\nu=2/3$ (blue circles). The magenta open circles are the measured noise along the upstream direction for shorter length ($4\,\mu$m) by reversing the magnetic field (opposite chirality), and injecting current at contact A and measured noise at contact B (see SM,S4). The solid black line is the theoretically calculated noise. (\textbf{e}) Noise measured along the downstream direction for $\nu=2/3$ for the scheme shown in Fig. 2b. 
}
\label{Figure2}
\end{figure}
%

\textbf{Experiment:} We turn now to a more detailed exposition of our results from two devices. The device schematic and noise measurement set-up is shown in Fig.~\ref{Figure1}c. The device fabrication details are mentioned in the method section and Supplementary Material (SM, S1). 
First, we perform electrical conductance measurements at a fixed magnetic field of $10$ T using standard lock-in amplifier. The electron density is tuned by the back graphite gate, keeping the top graphite gate fixed at zero voltage. The measured transverse Hall resistance ($R_{xy}$) and longitudinal resistance ($R_{xx}$) of device-1 are shown in Fig.~\ref{Figure1}d for $\nu\leq1$. 
Clear plateaus in $R_{xy}$ are developed for $\nu=1$, $\nu=2/3$, and $\nu=1/3$, accompanied with zeroes in $R_{xx}$ at the same gate voltage. Similarly, for $\nu=1$, $\nu=2/3$, $\nu=3/5$ and $\nu=1/3$ for device-2, see SM (S7).
Next, we perform noise measurements at $\nu=2/3$, $\nu=1/3$, and integer QH states for two different lengths $L = 10\,\mu$m and $4\,\mu$m of device-1. The Measurement scheme for $L = 10\,\mu$m is shown in Fig.~\ref{Figure2}a, where the device is set into $\nu=2/3$ QH state with $d$ and $u$ representing counter-propagating downstream and upstream eigenmodes. The measurement scheme for shorter length $L = 4\,\mu$m is obtained in the same device by changing the chirality (see SM, S4). Before the noise measurements, we perform two crucial checks: bias-dependent response of the $\nu=2/3$ FQH state and chirality of charge transport. For this, a 100\, pA AC signal is injected on top of the DC bias current at contact C, keeping contact D at ground, and the AC voltage at contacts C and B is measured. The voltage at contact C, shown as $V_{C}$ in Fig.~\ref{Figure2}c, remains flat as a function of the DC bias current, which means the conductance of the $\nu=2/3$ state does not change with the bias current. This 
kind of response, implying that there is no transport through the bulk (via states above the bulk gap),
is a prerequisite check for the noise measurements. This is in full consistency with the value of the gap of $\nu=2/3$ at 10\,T, which is around 5\,K as determined from the activation plot of $R_{xx}$ (see SM, S6), i.e., well above our largest bias voltage. Further, no detectable voltage is measured at contact B (shown as $V_{B}$ in Fig.~\ref{Figure2}c), which demonstrates that the charge propagates downstream only. Similar results are obtained for the shorter length ($4\,\mu$m) (see SM, S4). Thus, if our $\nu=2/3$ edge hosts counter-propagating modes (which is demonstrated below), the charge equilibration length is much smaller than the size of our device.

To unambiguously demonstrate the existence of upstream mode, a noiseless dc current is injected at contact C, and noise $S_I$ is measured at contact B along the upstream direction. For the noise measurement, a LCR resonant circuit with resonance frequency of $\sim$770\,kHz is utilized together with an amplifier chain and a spectrum analyzer (see methods and SM(S1))~\cite{srivastav2019universal,sahu2019enhanced,paul2020interplay,sahu2021quantized}. In the absence of an upstream mode, the energy cannot flow from the hot spot near C to the noise spot near B, see Fig.~\ref{Figure2}a, so that no noise is expected. This is precisely what is observed for $\nu=1$ and $\nu=1/3$ states, see Fig.~\ref{Figure2}d. At the same time, it is shown in Fig.~\ref{Figure2}d that for the $\nu=2/3$ state there is a strong noise which increases almost linearly with current. This is quite striking as at contact B the time-averaged current is zero. This clearly demonstrates that the $\nu=2/3$ edge hosts an upstream mode that leads to an upstream propagation of energy, even though the charge propagates downstream only. The mechanism of the noise generation is as follows \cite{spaanslatt2019topological}. The heat propagating upstream from the hot spot near C reaches the noise spot near B, inducing there creation of particle-hole pairs propagating in opposite directions. If the particle (or hole) is absorbed at contact B, while the hole (or, respectively, particle) flows downstream, there will be a voltage fluctuation at B detected by our noise measurement scheme. Similarly, the noise along the upstream direction is detected for $\nu=2/3$ and $\nu=3/5$ of device-2 as shown in SM(S7).

To verify that the heat propagates from the hot spot to the noise spot entirely via the edge, we have measured the noise in an alternative configuration. In this setting, a noiseless dc current is injected at contact A, while the contact C is electrically floating as shown in Fig.~\ref{Figure2}b. In this situation, in order to induce the noise, the heat would have to propagate upstream from the hot spot near D to the noise spot near B. However, this path is ``cut'' by the metallic contact C held at the base temperature. Thus, the only way for the heat to propagate is via the bulk. As shown in Fig.~\ref{Figure2}e, no detectable noise is measured at contact B, which rules out any sizeable bulk conduction of heat in our device.

\begin{figure}[!htb]
\centerline{\includegraphics[width=1.0\textwidth]{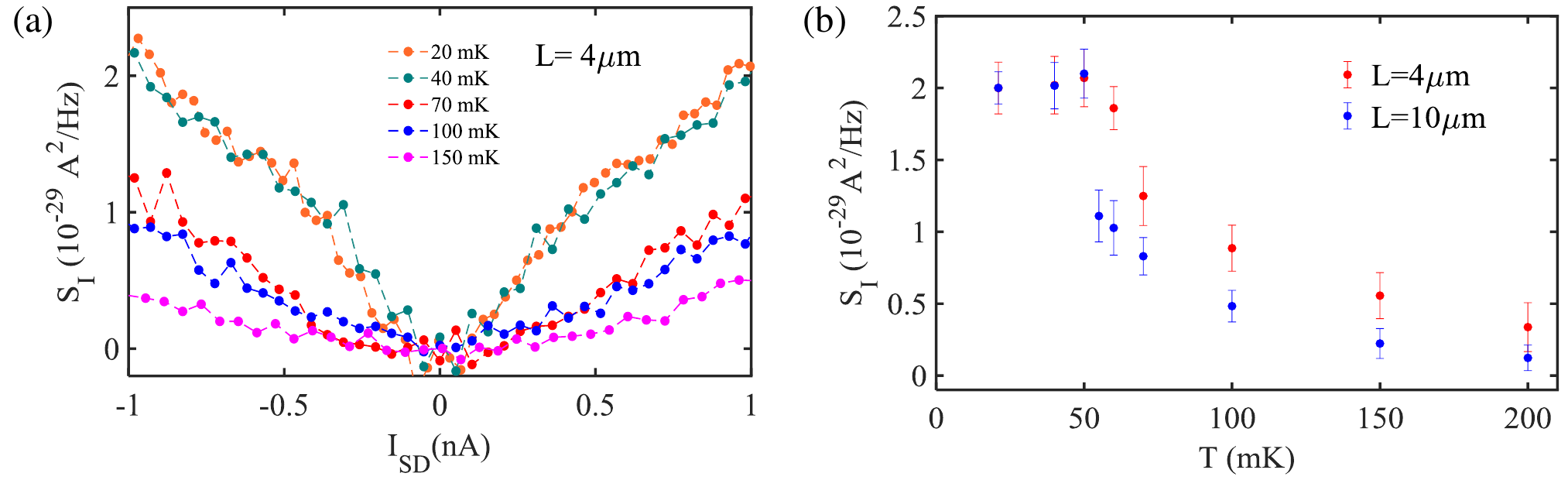}}
\caption{\textbf{Temperature dependence of the noise.} (\textbf{a}) Noise along the upstream direction at $\nu=2/3$ for different bath temperatures. (\textbf{b}) Noise value at $I_{S}=1$\,nA plotted against temperature for $4\,\mu$m (red circles) and $10\,\mu$m (blue circles) lengths. 
}
\label{Figure3}
\end{figure}
 
To inspect the length dependence of the noise, we have also studied it for $L= 4\,\mu$m of the same device-1 (see SM, S4). The data are shown in Fig.~\ref{Figure2}d. It can be seen that the noise amplitude is nearly identical for $L= 10\,\mu$m and $4\,\mu$m. This is striking since it shows that the heat propagates upstream ballistically and without losses along the $\nu=2/3$ edge, from the hot spot to the noise spot. There are two distinct mechanisms that could suppress the heat propagation and thus the noise: (i) thermal equilibration between the counter-propagating modes \cite{protopopov2017transport,nosiglia2018incoherent,Park2019noise,spaanslatt2019topological} and (ii) dissipation of energy from the edge to other degrees of freedom, including phonons, photons, and Coulomb-coupled localized states~\cite{schmidt2004photon,Gutman2016energy,rosenblatt2020energy}. Our results show that none of these mechanisms is operative at $T \sim 20$\,mK on the length scale of $L= 10\,\mu$m. The absence of thermal equilibration on the edge is in a striking contrast with the very efficient electric equilibration emphasized above. 
 
Up to now, all the data were at $T \sim 20$\,mK. Now, we explore the effect of temperature. Figure \ref{Figure3}a shows the evolution of the $\nu=2/3$ noise at $L= 4\,\mu$m with increasing $T$. In Fig.~\ref{Figure3}b we display the temperature dependence of the noise for $L= 4\,\mu$m and $L= 10\,\mu$m at $1nA$ current. It is seen that the noise remains constant and equal for both lengths up to $T = 50$\,mK. For higher temperature, the noise decays with $T$, and the decay is substantially faster for the larger $L$. This decay can be attributed to one of two mechanisms mentioned above: thermal equilibration within the edge (which would imply a crossover from ballistic to diffusive regime of heat flow) or loss of heat to the bulk. Further work is needed to understand which of them is dominant. 
 
Our experiments thus clearly indicate that at low temperatures, $T \le 50$\,mK, the upstream heat transport is ballistic and lossless. To support this conclusion, we have calculated theoretically the expected noise $S_I$ on the $\nu=2/3$ edge in this regime. The theory extends that of 
Refs.~\cite{Park2019noise,spaanslatt2019topological} to the ballistic (rather than diffusive) regime of heat transport corresponding to vanishing thermal equilibration. In this regime, the backscattering of heat takes place only at interfaces with the contact regions 
\cite{protopopov2017transport,melcer2021absent}. We assumed the bias voltage $V = (3h/2e^2) I_{SD}$ to be much larger than $T$, which is well fulfilled for our typical current $I_{SD} \sim 1nA$.  The result (see method section and SM (S9) for detail)
\begin{equation}
S_I = 0.146 \, e I_{SD} 
\end{equation}
is shown in Fig.~\ref{Figure2}d and is an excellent agreement with the experimental data, thus giving a further strong support to our interpretation of the experiment.

\textbf{Discussion:} Our measurements of noise present an unambiguous demonstration of the presence of an upstream mode in $\nu=2/3$ and $\nu=3/5$ FQH edges in graphene. This mode is responsible for the upstream heat transport that is at the heart of the noise generation mechanism. Remarkably, the noise is temperature-independent for $T \le 50$\,mK and remains the same for $L= 4\,\mu$m and $L= 10\,\mu$m, demonstrating the ballistic and lossless character of heat transport. The ballistic heat transport implies the absence of thermal equilibration on the edge, in 
contrast to 
full charge equilibration revealed by electric conductance measurements. This is entirely 
consistent with the data of Ref.~\cite{srivastav2021vanishing} on thermal conductance in graphene. There, a dramatic difference between the charge and heat equilibration lengths was explained by the vicinity of a system to a strong-interaction fixed point, where the bare modes of the $\nu=2/3$ edge are renormalized into a charge and a neutral mode. Very recently, absence of thermal equilibration (notwithstanding 
very efficient electric equilibration) was also reported for GaAs samples \cite{melcer2021absent}.

\newpage
\bibliography{ReferencesMS}
\pagebreak

\section{Methods}
\subsection{Device fabrication and measurement scheme:}
For making encapsulated devices, we used standard dry transfer pick-up technique~\cite{purdie2018cleaning,pizzocchero2016hot}. Fabrication of these heterostructures involved mechanical exfoliation of hBN and graphite crystals on oxidized silicon wafer using the widely used scotch-tape technique. BLG and graphite flakes were exfoliated from natural graphite crystals. Suitable flakes were identified under the optical microscope. The thickness of top and bottom graphite were $\sim 5\,$nm and $\sim 20\,$nm, respectively, and the thickness of of the top and bottom hBN flakes were of the order of $\sim 20\,$nm. The smaller distance between graphite gate and BLG layer ($\sim 20\,$nm thick bottom hBN) was similar range to the magnetic length scale of our experiment at $10$T, implying a sharp confining potential at the physical edge of BLG. Details of the fabrication procedure are in SM(S1). 
The BLG channel area of the stack was microscopically ironed using an AFM (atomic force microscopy) tip in contact mode~\cite{kim2019reliable}, to remove any atomic level strain or ripples or small bubbles from the channel area, which can arise due to stacking process. After this, for making contacts we used electron beam lithography (EBL). After EBL, reactive ion etching (mixture of CHF$_{3}$ and O$_{2}$ gas) 
was used to define the edge contact. 
Then, thermal deposition of Cr/Pd/Au (4/12/60\,nm) was done in a evaporator chamber having base pressure of $\sim$ $1-2\times 10^{-7}$ mbar. 
The optical image of the two measured devices are shown in SM(S1). The schematic of the device and measurement set-up are shown in Fig. 1(c). All the measurements are done in a cryo-free dilution refrigerator having base temperature of $\sim 20$mK. The electrical conductance was measured using standard Lock-in technique. For $R_{xy}$, an ac current is injected at contact C (Fig.~\ref{Figure1}c), contact A is kept at ground, and we measure the potential difference across the contacts D and B. For $R_{xx}$, we inject current at C, contact D is kept at ground, and we measure the potential difference across the contacts A and B. The resistance value of $R_{xy}$ for $\nu=2/3$ is $\approx 39k\Omega = (3/2)h/e^2 $, implying full charge equilibration among the counter-propagating eigenmodes. Under no charge equilibration, one would get very different value of the resistance according to Landauer-Buttiker formalism, see SM(S8) for detail. The noise is measured employing noise thermometry based on LCR resonant circuit at resonance frequency of $\sim 770$kHz and amplified by home made preamplifier at 4K followed by room temperature amplifier, and finally measured by a spectrum analyzer. The details of the noise measurement technique is mentioned in the SM(S1).

\subsection{Theoretical calculation of noise:}
We assume the regime of strong charge equilibration along the edge segment, i.e.,  $L\gg l^C_{\rm eq}$, where $l^C_{\rm eq}$ is the charge equilibration length. Then, the dc noise $S$ generated due to inter-mode tunneling along this segment is given by
 \begin{align}
 	S &=\frac{2e^2}{h l^C_{\rm eq}}\frac{ \nu_-}{ \nu_+} (\nu_+-\nu_-) \int_0^L dx\;\Lambda(x)e^{-\frac{2x}{l^C_{\rm eq}}}. \nonumber
  \end{align}
 Here, $h$ is the Planck constant, and $\nu_+$ and $\nu_-$ are the total filling factors associated with the downstream and upstream edge modes, respectively, with the bulk filling factor $\nu=\nu_+-\nu_-$. The exponential factor in the integral is a result of chiral charge transport, $\nu_+\neq \nu_-$, and implies that the dominant noise contribution comes from the noise spot---a region of size $\sim l^C_{\rm eq}$ near the contact that is located on the upstream side of the segment (the voltage probe where the noise is measured).  
 
 The noise kernel $\Lambda(x)$ is calculated by assuming a thermally non-equilibrated regime. In this case, $\Lambda$ becomes $x$-independent within the edge segment, so that the $x$-integral is straightforwardly calculated, yielding for the $\nu=2/3$ edge the noise 
 \begin{align}
	S_{2/3} = \frac{2e^2}{9h}\Lambda(V,\Delta) \,, \nonumber
\end{align}
where the dependence of $\Lambda$ on the bias voltage $V$ and the interaction between edge modes (parameter $\Delta$) is noted. We take $\Delta \approx 1$, which corresponds to the strong-interaction fixed point~\cite{kane1994randomness}, at which the heat equilibration length is much larger than the charge equilibration length~\cite{srivastav2021vanishing}, as observed experimentally. To determine the dependence of $\Lambda$ on the voltage, we first calculate effective temperatures of the downstream and upstream modes on the edge. They are found from the system of energy balance equations that include the Joule heating at the hot spot as well as a partial reflection of heat at interfaces between the interacting segment of the edge and the contact region. The contacts are modelled in terms of non-interacting 1 and 1/3 modes (see SM,S9). For $\Delta=1$, the corresponding reflection coefficient is $\mathcal{R}=1/3$. The resulting temperatures of the modes are
\begin{align}
k_BT_+ = 0.13 eV \,, \qquad \qquad k_BT_- = 0.23 eV \,.
\nonumber
\end{align}
The noise kernel $\Lambda$ is now calculated by using the Green's function formalism for the chiral Luttinger liquid and the Keldysh technique.  Expressing the result in terms of the bias current $I_{SD} = (2e^2/3h) V$, we come to the final result given by Eq.~(1) of the manuscript. An analogous calculation for $\nu=3/5$ edge yields $S_I = 0.138 \, eI_{SD}$, which is in a very good agreement with experiment too (see SM,S9).

\section{Acknowledgements}
C.S., Y.G., and A.D.M. acknowledge support by DFG Grants MI 658/10-1 and MI 658/10-2, and by the German-Israeli Foundation Grant No. I-1505-303.10/2019. Y.G. acknowledges support by the Helmholtz International Fellow Award, and by DFG RO 2247/11-1, and CRC 183 (project C01), and the Minerva Foundation. C.S. further acknowledges funding from the Excellence Initiative Nano at Chalmers University of Technology. K.W. and T.T. acknowledge support from the Elemental Strategy Initiative conducted by the MEXT, Japan and the CREST (JPMJCR15F3), JST. A.D. thanks the Department of Science and Technology (DST), India for financial support (DSTO-2051) and acknowledges the Swarnajayanti Fellowship of the DST/SJF/PSA-03/2018-19. R.K. and S.K.S acknowledge Inspire fellowship, DST and Prime Minister's Research Fellowship (PMRF), Ministry of Education (MOE), for financial support, respectively.


\section{Author contributions}
R.K. and S.K.S. contributed to device fabrication, data acquisition and analysis. A.D. contributed in conceiving the idea and designing the experiment, data interpretation and analysis. K.W and T.T synthesized the hBN single crystals. C.S., A.D.M. and Y.G. contributed in development of theory, data interpretation, and all the authors contributed in writing the manuscript.

\thispagestyle{empty}
\mbox{}


\section*{Supplementary material}

# Supplementary Material for "Observation of ballistic upstream modes at fractional quantum Hall edges of graphene"

Ravi Kumar[1], Saurabh Kumar Srivastav[1], Christian Spånslätt[2,3,4], K. Watanabe[5], T. Taniguchi[5], Yuval Gefen[6], Alexander D. Mirlin[3,4,7,8], and Anindya Das[1]

[1]*Department of Physics, Indian Institute of Science, Bangalore 560012, India*
[2]*Department of Microtechnology and Nanoscience,*
*Chalmers University of Technology, S-41296 Göteborg, Sweden*
[3]*Institut for Quantum Materials and Technology,*
*Karlsruhe Institute of Technology, 76021 Karlsruhe, Germany*
[4]*Institut für Theorie der Kondensierten Materie,*
*Karlsruhe Institute of Technology, 76128 Karlsruhe, Germany*
[5]*National Institute of Material Science, 1-1 Namiki, Tsukuba 305-0044, Japan*
[6]*Department of Condensed Matter Physics, Weizmann Institute of Science, Rehovot 76100, Israel*
[7]*Petersburg Nuclear Physics Institute, 188300 St. Petersburg, Russia and*
[8]*L. D. Landau Institute for Theoretical Physics RAS, 119334 Moscow, Russia*

PACS numbers:

This Supplementary Material contains the following details:

**S1.** Device Fabrication and noise measurement setup
**S2.** Gain calibration.
**S3.** Quantum Hall response
**S4.** Noise data for shorter length of the device
**S5.** Robustness of upstream noise across the $\nu = 2/3$ plateau
**S6.** Energy gap of FQH states.
**S7.** Device-2 data for $\nu = 2/3$ and $\nu = 3/5$ states
**S8.** Electrical conductance for no charge equilibration using Landauer-Büttiker model
**S9.** Theoretical calculation of noise for $\nu = 2/3$ and $\nu = 3/5$ states

## S1. DEVICE FABRICATION AND NOISE MEASUREMENT SETUP

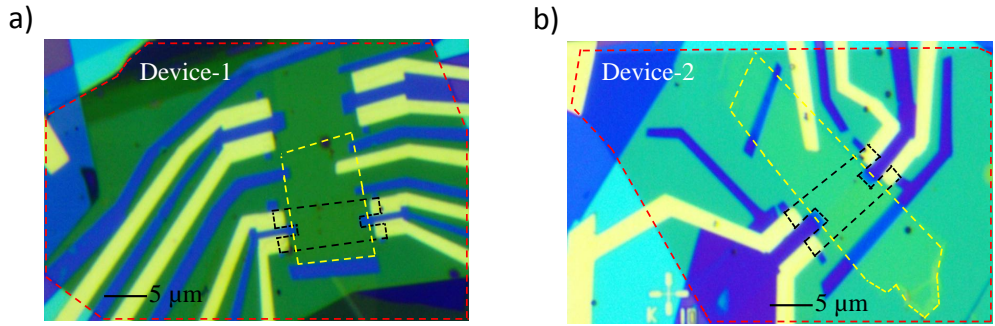

FIG. S1: **Optical image of the devices.** (**a**) and (**b**) show optical images of devices 1 and 2, respectively. Region of BLG, top graphite, and bottom graphite are marked by black, yellow, and red dashed lines, respectively.

For making encapsulated devices (heterostructure of graphite/hBN/bilayer graphene(BLG)/hBN/graphite), we used standard dry transfer pick-up technique [S1, S2]. Fabrication of these heterostructure involved mechanical exfoliation of hBN and graphite crystals on oxidized silicon wafer using the widely used scotch-tape technique. BLG and graphite flakes were exfoliated from natural graphite crystals. Suitable flakes were identified under the optical microscope. First, a graphite of thickness $\sim 5\,\text{nm}$ which works as top gate or screening layer, is picked up at 90°C using a Poly-Bisphenol-A-Carbonate (PC) coated Polydimethylsiloxane (PDMS) stamp placed on a glass slide, attached to tip of a home-built micromanipulator. Subsequently, this graphite flake is aligned on top of an already exfoliated hBN flake of thickness $\sim 20\,\text{nm}$ at 90°C, and due to better adhesion of hBN with graphite than $SiO_2$, hBN flake gets picked up. Next, we align this graphite/hBN stamp on top of a BLG flake. BLG was picked up at the same 90°C temperature. Next step involved the pick-up of bottom hBN ($\sim 20\,\text{nm}$). Bottom hBN was picked up using the previously picked-up graphite/hBN/BLG following the previous process. This graphite/hBN/BLG/hBN heterostructure was used to pick up the bottom graphite flake following previous step. Finally, this resulting heterostructure (graphite/hBN/BLG/hBN/graphite) was dropped down on top of a silicon wafer with 285 nm thick $SiO_2$ on top, at temperature 180°C. To remove the residues of PC, the final stack was cleaned in chloroform ($CHCl_3$) overnight followed by cleaning in acetone and and iso-propyl alcohol (IPA). The BLG channel area of the stack was microscopically ironed using an AFM (atomic force microscopy) tip in contact mode [S3], to remove any atomic level strain or ripples or small bubbles from the channel area, which can arise due to stacking process. After this, Poly-methyl-methacrylate (PMMA) photoresist was coated on this heterostructure to define the contacts using electron beam lithography (EBL). After EBL, reactive ion etching (mixture of $CHF_3$ and $O_2$ gas with flow rate of 40 sccm and 4 sccm, respectively, at 25°C with RF power of 60W) was used to define the edge contact. The etching time was optimized such that the bottom hBN does not get etched out completely to isolate the contacts from bottom graphite flake, which was used as the back gate. Then, thermal deposition of Cr/Pd/Au (4/12/60 nm) was done in a evaporator chamber having base pressure of $\sim 1 - 2 \times 10^{-7}$ mbar. After deposition, lift-off procedure was performed in hot acetone and IPA. Finally, to define the BLG edge along the shorter length, we again performed reactive ion etching. The optical image of the devices and of the measurement setup are shown in Figures S1 and S2, respectively.

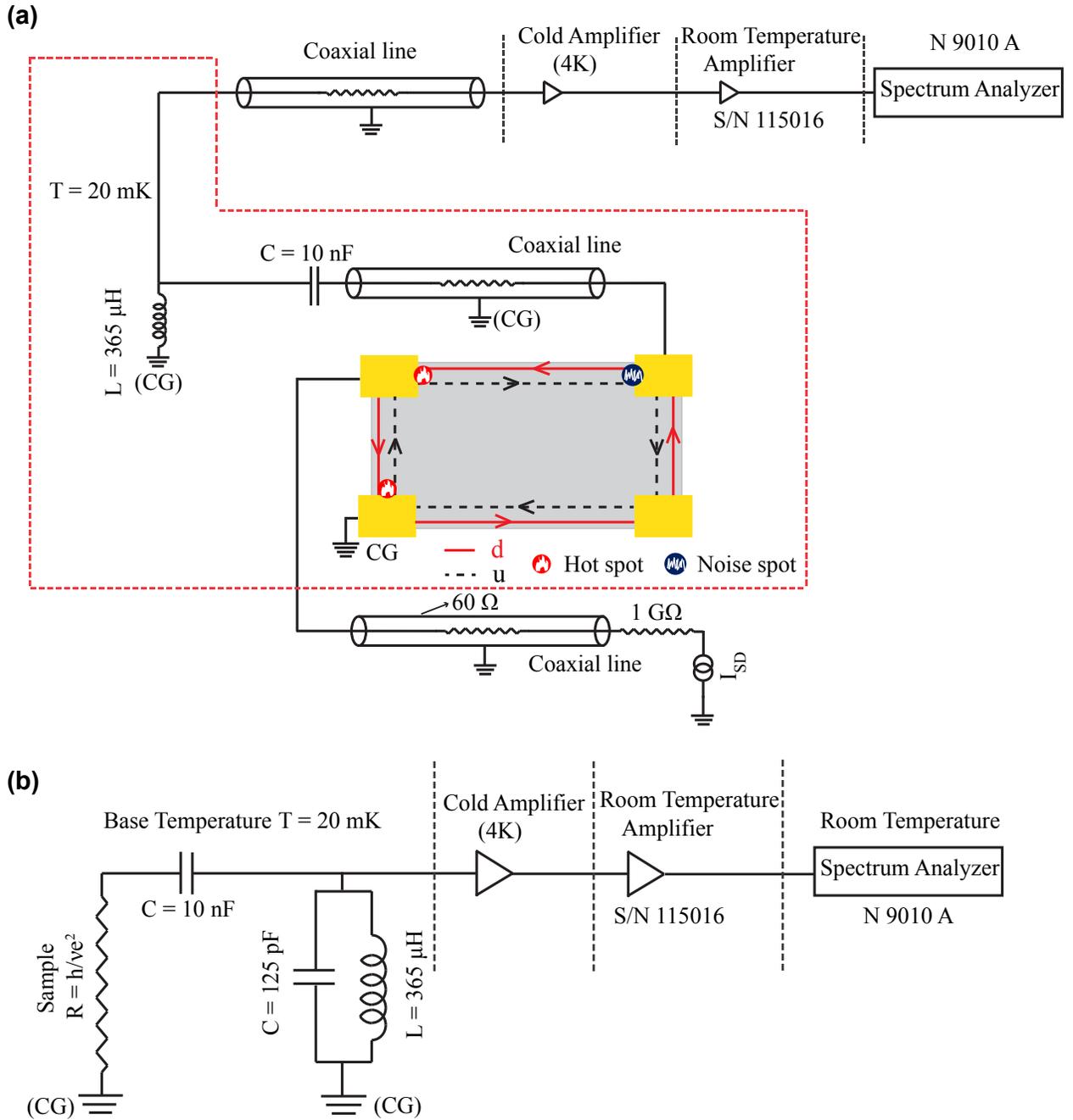

FIG. S2: **Experimental setup for noise measurement.** (a) Schematic of the measurement setup. The device was mounted on a chip carrier which was connected to the cold finger fixed to the mixing chamber plate of dilution refrigerator. The ground contact pins are directly shorted to the cold finger to achieve the cold ground. The sample was current biased with a 1 G$\Omega$ resistor located at the top of dilution fridge. Current fluctuations measured at contact located along the upstream direction are converted on chip into voltage fluctuations using the quantum Hall (QH) resistance $R = h/\nu e^2$ [as shown schematically in figure (b)] where $\nu$ is the filling factor. The voltage noise generated from the device is filtered using a superconducting resonant LC tank circuit with resonance frequency of 770kHz and bandwidth 30kHz. The filtered signal is then amplified by a home made cryogenic voltage pre-amplifier which was thermalized to 4K plate of dilution refrigerator. This pre-amplified signal was then amplified using a voltage amplifier (PR-E3-SMA S/N 115016) placed at the top of the fridge at room temperature. After second stage of amplification, amplified signal was measured using a spectrum analyzer (N9010A). All the noise measurements were done using the band width of $\sim$ 30 kHz. The resonant LC tank circuit was built using inductor L of $\sim$ 365 $\mu$H made from a superconducting coil thermally anchored to the mixing chamber of dilution refrigerator. The parallel C of $\sim$ 125 pF is the capacitance that develops along the coaxial lines connecting the sample to the cryogenic pre-amplifier. A ceramic capacitance of 10 nF was introduced between sample and inductor to block the DC current along the measurement line. The typical input voltage noise and current noise of cryogenic pre-amplifiers were $\sim$ 250 pV/$\sqrt{\text{Hz}}$ and $\sim$ 20-25 fA/$\sqrt{\text{Hz}}$, respectively. (b) The schematic circuit diagram of the measurement setup.

## S2. GAIN CALIBRATION

We have estimated the gain of amplification chain from temperature-dependent Johnson-Nyquist noise (thermal noise). At zero impinging current, the equilibrium voltage noise spectrum is given by

$$S_V = g^2(4k_B T R + V_n^2 + i_n^2 R^2)BW,  \tag{S1}$$

where $g$ is the total gain of amplification chain, $k_B$ the Boltzmann factor, $T$ the temperature, $V_n^2$ and $i_n^2$ are the intrinsic voltage and current noise of the amplifier, and $BW$ is the frequency bandwidth. At an integer quantum Hall plateau, any change in temperature of mixing chamber (MC) plate will only affect the first term in Eq. (S1), while all other terms are independent of temperature. If one plot the $\frac{S_V}{BW}$ as a function of temperature, the slope of the linear curve will be equal to $4g^2 k_B R$. Since at quantum Hall plateau resistance R is exactly known, one can easily extract the gain of the amplification chain from the slope and the intrinsic noise of the amplifier from the intercept. Thus, the gain g is found using the following equation

$$g = \sqrt{\Big(\frac{\partial\big(\frac{S_V}{BW}\big)}{\partial T}\Big)\Big(\frac{1}{4k_B R}\Big)},  \tag{S2}$$

where $\Big(\frac{\partial\big(\frac{S_V}{BW}\big)}{\partial T}\Big)$ is the slope of the linear fit. The implementation of this procedure is shown in Fig. S2.

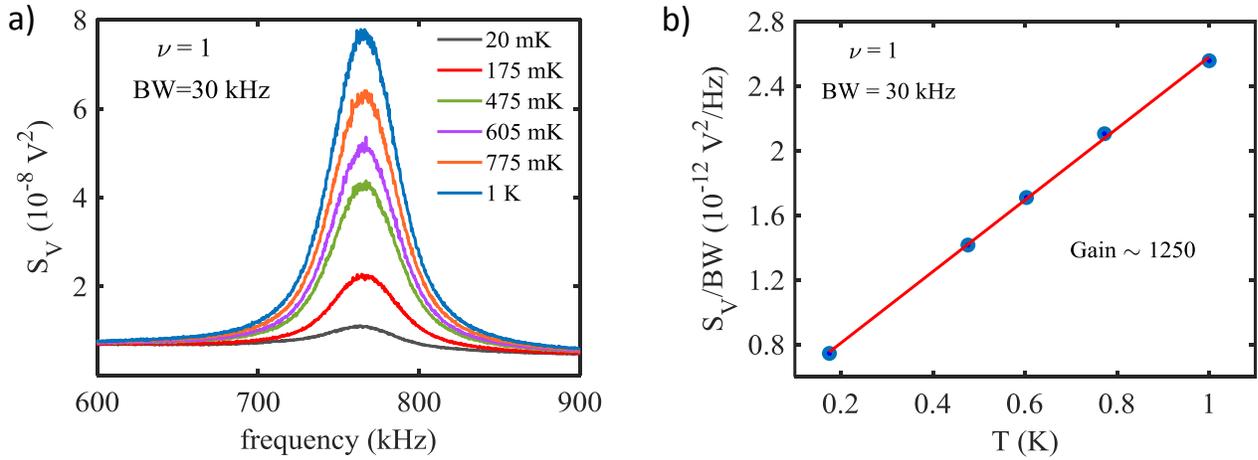

FIG. S3: **Gain of the amplification chain.** (**a**) Noise $S_V$ measured by spectrum analyzer is plotted as a function of frequency at different temperature for $\nu = 1$. From this plot, resonance frequency of tank circuit was found to be ∼763 kHz. (**b**) Symbols represent the noise $S_V$ divided by bandwidth ($BW$) at resonance frequency as a function of temperature. Solid red line is the linear fit of data. Using Eq. (S2) and the slope of this linear fit, the gain $g$ was found to be equal to ∼1250.

## S3. QUANTUM HALL CHARACTERIZATION

We characterize QH response of the device-1 for both the chiralities of QH edge and for different edge length of the device as used in the noise measurements. Details are shown in Fig. S4.

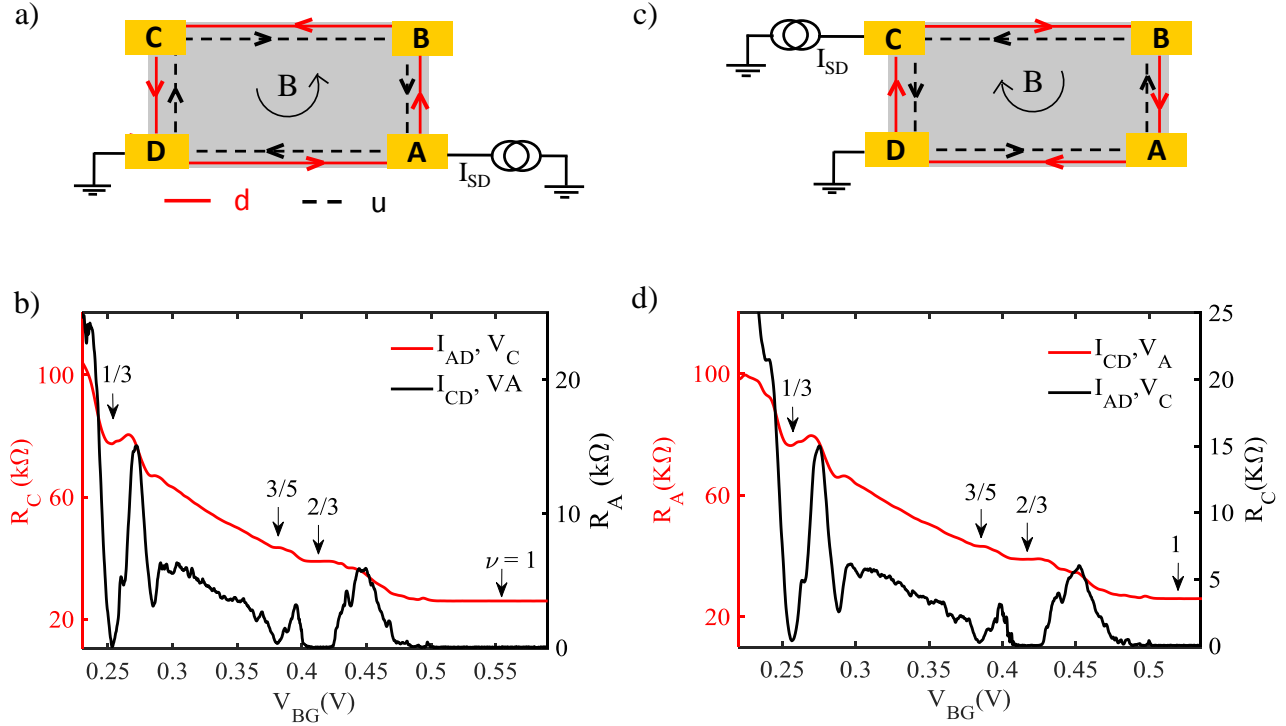

FIG. S4: (**a**) Measurement scheme for checking low-frequency QH response for anticlockwise chirality (+10 T). An ac current $I_{SD}$ of 2 nA magnitude at frequency $f = 13$ Hz from Lock-In is injected at contact A, contact D is cold ground, and voltage is measured at contact C. The measured voltage is converted to resistance, shown as $R_C$ (red color) in panel (b). Similarly, 2 nA ac current at 13 Hz is injected at contact C and resistance is measured at contact A, shown as $R_A$ (black color) in panel (b). We see clear quantized plateaus in $R_C$ and vanishing resistance in $R_A$ at $\nu = 1$, 2/3, and 1/3. (**c**) Measurement scheme for checking low frequency QH response for clockwise chirality( −10 T). An ac current $I_{SD}$ of 2 nA magnitude at frequency $f = 13$ Hz from Lock-In is injected at contact C, contact D is cold ground, and resistance is measured at contact A, shown as $R_A$ (red color) in panel (d). Similarly, 2 nA ac current at 13 Hz is injected at contact A and resistance is measured at contact C, shown as $R_C$ (black color) in panel (d). We see clear quantized plateaus in $R_A$ and vanishing resistance in $R_C$ at $\nu = 1$, 2/3, and 1/3.

## S4. NOISE DATA FOR SHORTER LENGTH

To perform noise measurements along the shorter length of the device-1, we reversed the chirality of the QH edge by changing the magnetic field from $+10\,\text{T}$ to $-10\,\text{T}$. Measurement scheme for this is shown in panel (**a**). Before noise measurements, we again check the linearity of the bias response of the $\nu = 2/3$ state as well as the downstream character of the charge propagation by injecting a 100 pA AC signal from Lock-In on top of the DC bias current at contact A, contact D is cold grounded, and voltage is measured at contacts A and B, as shown in panel (**e**), cf. Fig. 2c of the main manuscript. Flat $V_A$ indicates that the conductance of $\nu = 2/3$ state is independent of bias current. Flat zero value of $V_B$ shows that the charge propagates entirely downstream and only energy can propagate upstream. To detect the upstream heat transport, a noiseless dc current is injected at contact A, and noise is measured at contact B along the upstream direction. Due to downstream current $I_A$, a hot spot is created at the backside of the contact A. Part of this heat is carried by the upstream mode towards the noise spot near the contact B, thus generating noise at contact B. The measured noise for $\nu = 2/3$ and $\nu = 1$ is shown in the panel (**b**). For $\nu = 1$, no noise is detected, in full consistency with the absence of upstream modes at this filling. For $\nu = 2/3$, the noise is nearly identical to that for $L = 10\,\mu\text{m}$, see Fig. 2d of the main manuscript. In panel (**c**), an alternative configuration is shown: we inject noiseless dc current at contact C and measure noise at contact B. The contact A is electrically floating, thus cutting the path along the edge from the hot spot at D to the noise spot at B. No noise is detected for $\nu = 2/3$ state, see panel (**d**), confirming that the heat propagation responsible for the noise generation takes place along the edge, cf. Fig. 2e of the main manuscript.

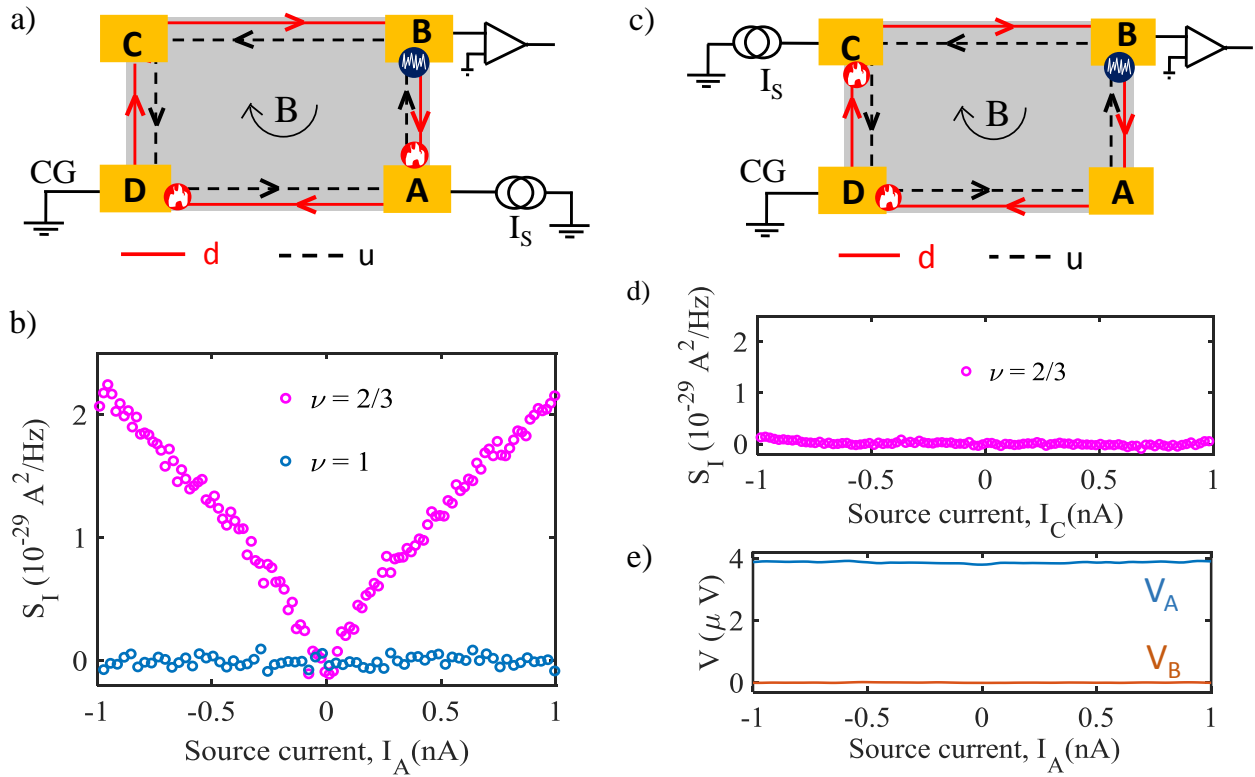

FIG. S5: (**a**) Device measurement scheme for upstream noise detection along the shorter length ($4\mu m$) of the device. Device is set into $\nu = 2/3$ state where d and u represent counter-propagating eigen-modes. (**b**) Upstream noise for hole-conjugate state $\nu = 2/3$ and electron-like state $\nu = 1$, measured along the shorter length. (**c**) and (**d**) represent measurement scheme and the measured noise at 2/3 for an alternate contacts configuration, respectively. (**e**) Bias response of 2/3 state.

## S5. ROBUSTNESS OF NOISE ACROSS THE $\nu = 2/3$ PLATEAU

The noise at the $\nu = 2/3$ state is measured at different gate voltages inside the 2/3 plateau for both chiralities of the QH edge. The noise remains same, as shown in Fig.S6.

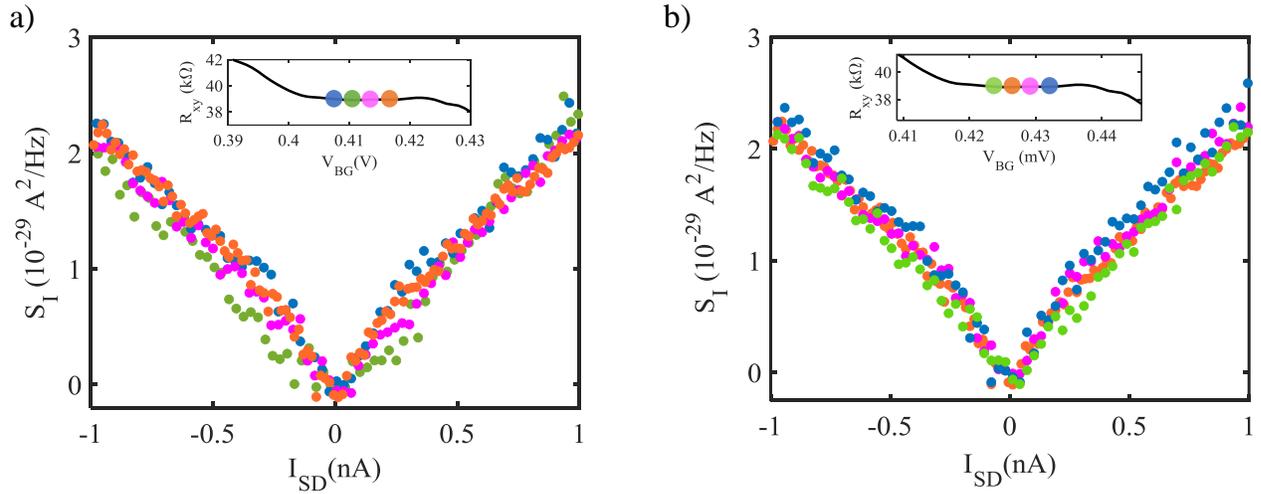

FIG. S6: (a) and (b) show the noise measured along the longer length ($10\,\mu$m) and shorter length ($4\,\mu$m) of the device, respectively, at different values of the density (controlled by the back gate voltage $V_{BG}$) on the $\nu = 2/3$ plateau. Insets show the Hall-resistance plateau of the $\nu = 2/3$ state. Different colors in noise data correspond to different values of the gate voltage inside the plateau, as marked by the corresponding color dots in the insets.

## S6. ENERGY GAP OF FQH STATES

Here (Fig. S7) we present experimental determination of the energy gap of $\nu = 1/3$ and $\nu = 2/3$ states by temperature activated behavior of the longitudinal resistance $R_{xx}$ at the fixed magnetic field 10 T.

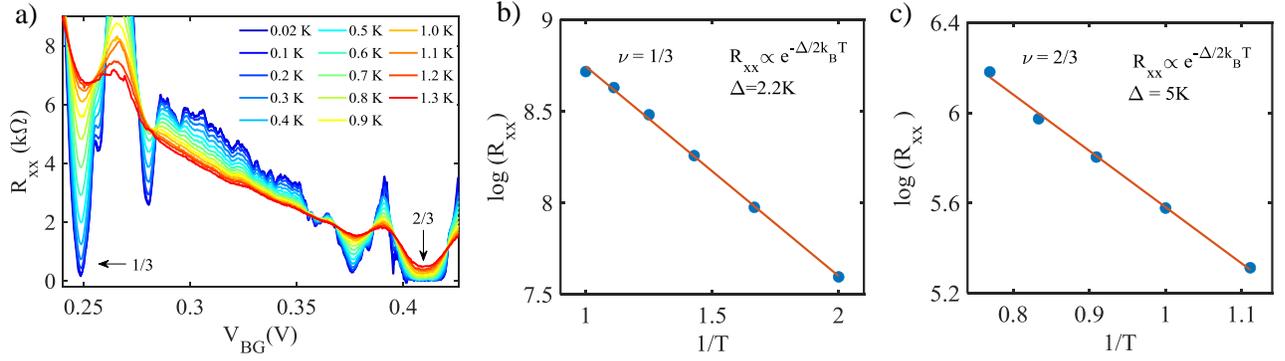

FIG. S7: **Activation energy gap of $\nu = 1/3$ and $\nu = 2/3$ state.** (a) Gate dependence of the longitudinal resistance $R_{xx}$ measured at 10 T magnetic field at different temperatures, starting from base temperature of 20 mK up to 1.3 K. (b) Logarithm of $R_{xx}$ vs inverse temperature $1/T$ for $\nu = 1/3$ state. The linear fit yields $R \propto \exp(-\Delta/2k_B T)$ with $\Delta = 2.2$ K. (c) Logarithm of $R_{xx}$ vs inverse temperature $1/T$ for $\nu = 2/3$ state. The linear fit yields $R \propto \exp(-\Delta/2k_B T)$ with $\Delta = 5$ K.

## S7. DEVICE-2 DATA

The noise measurements for the $\nu = 2/3$ state were also carried out in device-2 (shown in Fig. S1b) for $10\,\mu$m propagation length. Furthermore, we were able to measure in device-2 noise for the $\nu = 3/5$ state too, due to a better longitudinal response (more pronounced quantum Hall plateau) of 3/5 state in device-2. Data is shown in Fig. S8. The results for the $\nu = 2/3$ state (Fig. S8b) are in full agreement with those for device-1. The results for the $\nu = 3/5$ state (Fig. S8c) confirm the existence of upstream modes for this state as well. The magnitude of the noise for the $\nu = 3/5$ state is rather close to that for the $\nu = 2/3$ state and is in a very good agreement with the theoretical prediction.

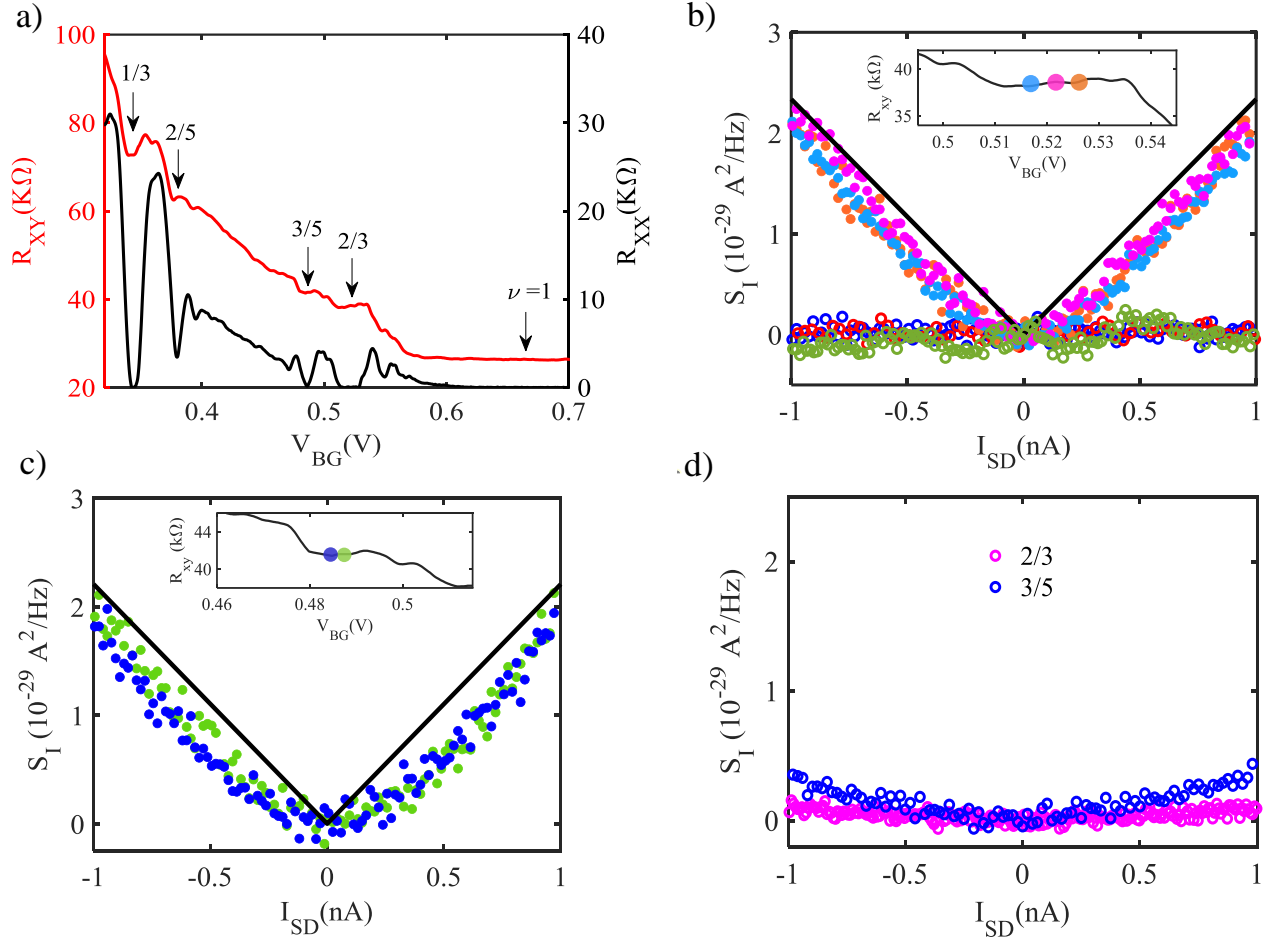

FIG. S8: **Device-2 data.** (**a**) Quantum Hall response of device-2. (**b**) Noise data for $\nu = 2/3$, 1, and 2. Solid circles correspond to noise for 2/3, with different color data corresponding to different gate voltage position inside the 2/3 plateau as shown in the inset. The value of the noise perfectly agrees with that for device-1, see Fig. 2d of the main manuscript, and with the theoretical prediction (solid black line) $S_I = 0.146\, I_{\rm SD}$, see Eq. (S41) below. Open olive, blue and red circles correspond to the noise for $\nu = 1/3$, 1 and 2, respectively. No noise is detected, in full agreement with the absence of upstream modes for these fillings. (**c**) Noise for $\nu = 3/5$. The magnitude of the noise agrees very well with the theoretical prediction (solid black line), $S_I = 0.138\, I_{\rm SD}$, see Eq. (S54) below. (**d**) Noise for $\nu = 2/3$ and 3/5 in alternative configuration of contacts, with the upstream path from hot spot to the noise spot cut by a floating metallic contact. For $\nu = 2/3$, no noise is detected, as in device-1, demonstrating that the heat propagation from hot spot to noise spot, which is responsible for the noise generation, takes place along the edge. For $\nu = 3/5$ a very weak noise is detected, which is apparently due to a contribution of heat transport through the bulk, possibly via the mechanism identified in Ref. [S4]. While this contribution is totally negligible for the $\nu = 2/3$ state, it becomes detectable for the $\nu = 3/5$ in view of a smaller value of the gap.

## S8. ELECTRICAL CONDUCTANCE IN THE ABSENCE OF CHARGE EQUILIBRATION USING LANDAUER-BÜTTIKER FORMALISM

As demonstrated in the main text, the charge propagates only downstream in our devices, which implies that the charge equilibration length is much smaller than the relevant propagation length $L$ ($4\,\mu$m or $10\,\mu$m). Here we provide further evidence of strong charge equilibration in our devices. For this purpose, we calculate resistances assuming no charge equilibration in two configuration of contacts. We show that these values are very different from experimentally observed resistances that, on the other hand, are in full agreement with the values predicted for a regime of full charge equilibration.

To calculate the value of electrical conductance of the $\nu = 2/3$ state without charge equilibration, we follow the approach of Landauer-Büttiker model [S5]. Note that in this calculation we assume full equilibration at the contacts. We calculate the electrical conductance of $2/3$ state, which host counter propagating edge states of conductance $1e^2/h$ and $(1/3)e^2/h$. The schematic of the device with two different contact configurations is shown in Fig. S9a,b.

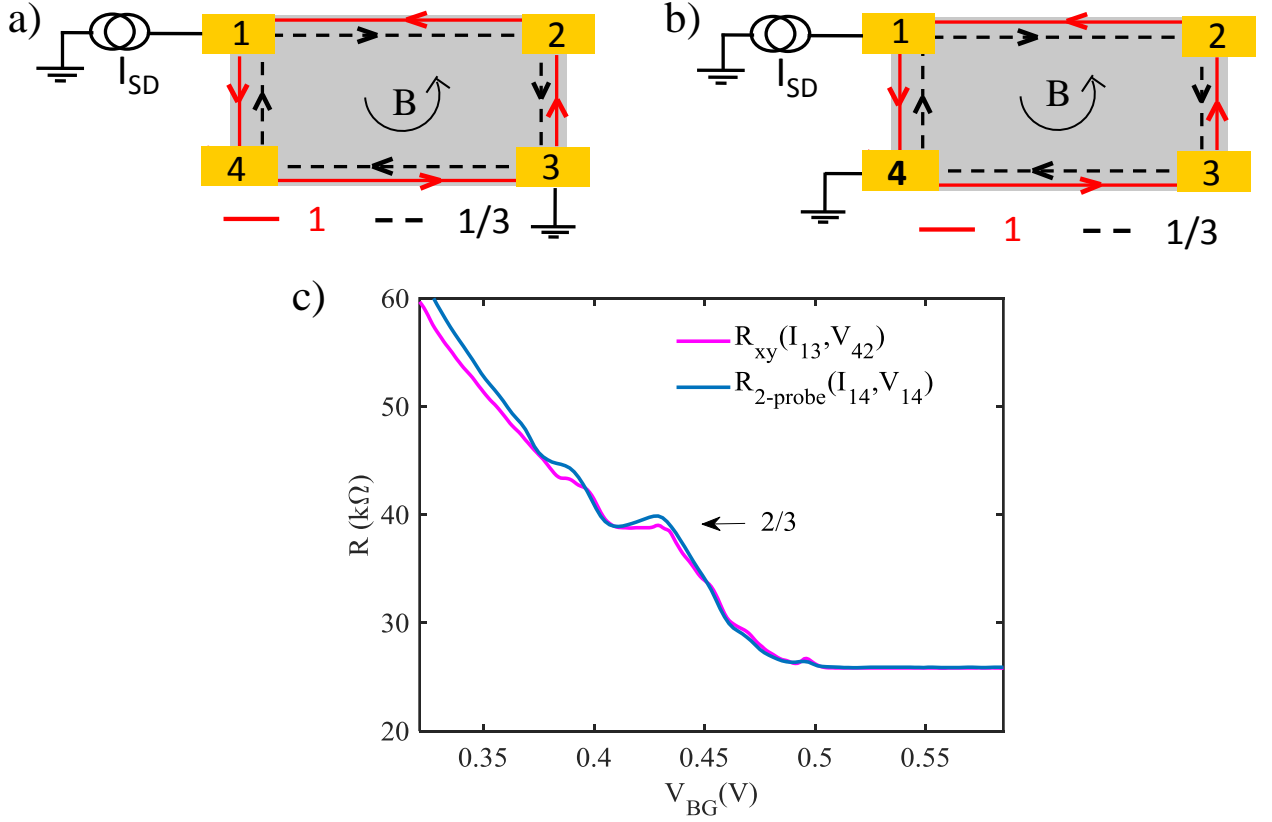

FIG. S9: (**a**) Contact configurations to measure transverse Hall resistance of $2/3$ state. The contact 1 is biased, 3 is grounded, 2 and 4 floating. (**b**) An alternative contact configuration: 1 is biased, 4 is grounded, 2 and 3 floating. (**c**) Measured transverse Hall resistance $R_{xy}$ (magenta color) for the first contact configuration and resistance at contact 1 (blue color) for the second contact configuration. Legend notations: the first index in the subscript of $I$ corresponds to the current-fed contact and the second one to the ground contact. Indices in the subscript of $V$ correspond to contacts across which the potential difference is measured.

For a multi-probe device, the net current flowing in $i^{th}$ contact is given by

$$I_i = \sum_j (G_{j\leftarrow i} V_i - G_{i\leftarrow j} V_j), \tag{S3}$$

where $G_j \leftarrow i$ is the conductance from $i^{th}$ contact to $j^{th}$ contact and $V_i$ is the voltage of $i^{th}$ contact. In the absence

of charge equilibration along the edges, we would have

$$\begin{pmatrix} I_1 \\ I_2 \\ I_3 \\ I_4 \end{pmatrix} = \frac{e^2}{h} \begin{pmatrix} (1+1/3) & -1 & 0 & -1/3 \\ -1/3 & (1+1/3) & -1 & 0 \\ 0 & -1/3 & 1+1/3 & -1 \\ -1 & 0 & -1/3 & (1+1/3) \end{pmatrix} \begin{pmatrix} V_1 \\ V_2 \\ V_3 \\ V_4 \end{pmatrix}. \tag{S4}$$

Eliminating the rows and column associated with contact 3 from Eq. (S4), we get

$$\begin{pmatrix} I_1 \\ I_2 \\ I_4 \end{pmatrix} = \frac{e^2}{h} \begin{pmatrix} 4/3 & -1 & -1/3 \\ -1/3 & 4/3 & 0 \\ -1 & 0 & 4/3 \end{pmatrix} \begin{pmatrix} V_1 \\ V_2 \\ V_4 \end{pmatrix}. \tag{S5}$$

The current is injected at contact 1, so that

$$\begin{pmatrix} I_1 \\ I_2 \\ I_4 \end{pmatrix} = \begin{pmatrix} I \\ 0 \\ 0 \end{pmatrix}. \tag{S6}$$

According to Eq. (S5), this yields the voltages

$$\begin{pmatrix} V_1 \\ V_2 \\ V_4 \end{pmatrix} = I \frac{h}{e^2} \begin{pmatrix} 1.2 \\ 0.3 \\ 0.9 \end{pmatrix}, \tag{S7}$$

so that the voltages measured at contacts 1, 2, and 4 will be

$$V_1 = 1.2 \times I \frac{h}{e^2}, \qquad V_2 = 0.3 \times I \frac{h}{e^2}, \qquad V_4 = 0.9 \times I \frac{h}{e^2}. \tag{S8}$$

Hence the transverse Hall resistance in the absence of charge equilibration is given by

$$R_{xy} = \frac{V_4 - V_2}{I} = 0.6 \times \frac{h}{e^2} = 15.5 \, \text{k}\Omega. \tag{S9}$$

At the same time, the experimentally measured value of $R_{xy}$ is $\approx 38.6 k\Omega$, as shown in Fig. S9(c) in magenta color.

Next we consider the configuration shown in Fig. S9(b). Here the contact 4 is grounded. After eliminating the rows and column associated with contact 4 from the matrix Eq. (S4), we get

$$\begin{pmatrix} I_1 \\ I_2 \\ I_3 \end{pmatrix} = \frac{e^2}{h} \begin{pmatrix} 4/3 & -1 & 0 \\ -1/3 & 4/3 & -1 \\ 0 & -1/3 & 4/3 \end{pmatrix} \begin{pmatrix} V_1 \\ V_2 \\ V_3 \end{pmatrix}. \tag{S10}$$

The current is injected at contact 1 also in this configuration, so that

$$\begin{pmatrix} I_1 \\ I_2 \\ I_3 \end{pmatrix} = \begin{pmatrix} I \\ 0 \\ 0 \end{pmatrix}. \tag{S11}$$

Hence, the voltages are

$$\begin{pmatrix} V_1 \\ V_2 \\ V_3 \end{pmatrix} = I \frac{h}{e^2} \begin{pmatrix} 0.975 \\ 0.300 \\ 0.075 \end{pmatrix}, \tag{S12}$$

so that the voltages measured at contact 1 will be

$$V_1 = 0.975 \times I \frac{h}{e^2}. \tag{S13}$$

Thus, the two-terminal resistance between the contacts 1 and 4

$$R_{2-\text{probe}, \, 14} = \frac{V_1}{I} = 0.975 \times \frac{h}{e^2} = 25.17 \, \text{k}\Omega. \tag{S14}$$

At the same time, the experimentally measured value of the resistance between the contacts 1 and 4 is $\approx 38.6\,\text{k}\Omega$.

Thus, experimentally measured values of resistances in both configurations are very different from those that one would have in the absence of charge equilibration between counter-propagating modes in the 2/3 edge. At the same time, these values correspond exactly to the limit of strong charge equilibration between the modes. Indeed, in this limit, the edge can be thought of (in the sense of charge transport) as a single mode with the conductance $(2/3)e^2/h$, which yields the resistance $(3/2)h/e^2$ for both configurations—which is exactly the measured values. This confirms that charge equilibration is fully developed for both propagation lengths in our device.

## S9. THEORETICAL CALCULATIONS OF NOISE

### A. Preliminaries

We consider a FQH edge segment with length $L$ connected to two contacts, as depicted in Fig. S10. Our model of this segment consists of three regions: two non-interacting contact regions and one central, interacting region. We are interested in noise in the large bias regime $eV_0 \gg k_B T_0$, where $e$ is the elementary charge, $V_0$ is the bias voltage, $k_B$ is the Boltzmann constant and $T_0$ is the system base temperature. Under this condition, we set $T_0 = 0$ in the following. We assume the regime of strong charge equilibration along the edge segment, i.e., $L \gg l_{\rm eq}^C$, where $l_{\rm eq}^C$ is the charge equilibration length. Then, the dc noise $S$ generated due to inter-mode tunneling along this this segment can be written as [S6–S9]

$$S = \frac{2e^2}{h l_{\rm eq}^C} \frac{\nu_-}{\nu_+} (\nu_+ - \nu_-) \int_0^L dx\, \Lambda(x) e^{-\frac{2x}{l_{\rm eq}^C}}. \tag{S15}$$

Here, $h$ is the Planck constant, and $\nu_+$ and $\nu_-$ are the total filling factors of the downstream and upstream edge modes, respectively, with the bulk filling factor $\nu = \nu_+ - \nu_-$. The exponential factor in the integral is a result of chiral charge transport, $\nu_+ \neq \nu_-$, and implies that the dominant noise contribution comes from a region of size $\sim l_{\rm eq}^C$ close to the left contact, the so-called noise spot. In Eq. (S15), we have neglected thermal fluctuations emanating from the contacts, as these fluctuations are much weaker than the non-equilibrium fluctuations induced by the bias, in view of the condition $eV_0 \gg k_B T_0$.

The main quantity to compute in Eq. (S15) is the local noise kernel

$$\Lambda(x) = \frac{S_{\rm loc}[\Delta V(x), T_+(x), T_-(x)]}{2 g_{\rm loc}[\Delta V(x), T_+(x), T_-(x)]}, \tag{S16}$$

where $S_{\rm loc}$ and $g_{\rm loc}$ are the local electron-tunneling dc noise and the tunneling conductance, respectively. These quantities depend on microscopic details of the edge, including the inter-mode interactions, the local voltage difference between the modes $\Delta V(x)$, and the effective temperatures $T_\pm(x)$ of downstream $(+)$ and upstream $(-)$ edge modes. Below we compute $\Lambda(x)$, and the resulting noise $S$, for the edges at fillings $\nu = 2/3$ (in Sec. S9 B) and $\nu = 3/5$ (in Sec. S9 C).

Note that the previous works [S6–S9] where the theory of noise on a FQH edge was developed assumed that charge (subscript "C") and heat (subscript "H") equilibration lengths are of the same order, $l_{\rm eq}^H \sim l_{\rm eq}^C$. The focus there was on the regime of strong charge and heat equilibration, $L \gg l_{\rm eq}^H \sim l_{\rm eq}^C$. On the other hand, experimental results of Ref. [S10] (where the thermal conductance was studied) and of the present work on graphene samples show that the system is in the regime $l_{\rm eq}^C \ll L \ll l_{\rm eq}^H$ and thus the two equilibration length differ very strongly. Emergence of this regime was explained theoretically in Ref. [S10] as the effect of a strong inter-mode interaction. Specifically, it was shown there that, when a parameter $\Delta$ characterizing the interaction strength (see below for more detail) approaches unity, the

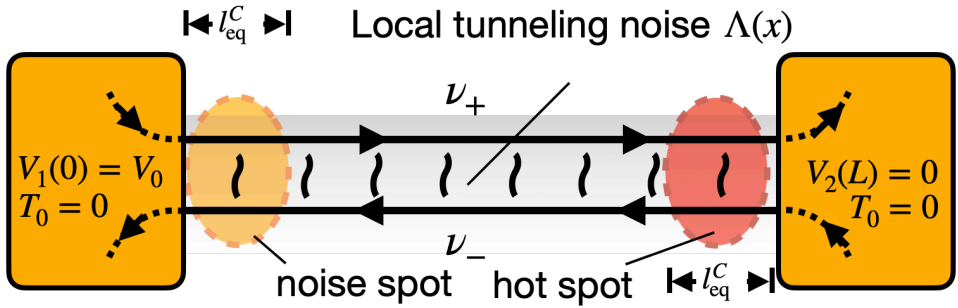

FIG. S10: Noise generation on the edge. The two contacts are biased with the voltage $V_0$ satisfying $eV_0 \gg k_B T_0$. The voltage drop and Joule heating occurs in a region of size $\sim l_{\rm eq}^C$ close to the right contact. This is called the hot spot. Excess dc noise is predominantly generated instead in the region of the spatial extent $\sim l_{\rm eq}^C$ close to the left contact. This is the noise spot. The generated noise is described by Eq. (S15). In the experiment, the generated noise is found by measuring the voltage fluctuations of the left contact (which is on the upstream side of the segment) that is floating and serves as a voltage probe. Left and right contacts of this figure correspond to contacts B and C in Fig. 2a of the main manuscript, respectively.

thermal equilibrium diverges as $l_{\text{eq}}^H \sim (\Delta - 1)^{-1}$, while $l_{\text{eq}}^C$ does not show this singularity. Very recently, the regime $l_{\text{eq}}^C \ll L \ll l_{\text{eq}}^H$ was also observed in GaAs structures [S11]. Our main goal here is the calculation of the noise (S15) for $\nu = 2/3$ and $\nu = 3/5$ edges in this regime of strong charge equilibration but vanishing thermal equilibration. Parallel to this work, a related calculation of the so-called $\Delta T$-noise (i.e., induced by a temperature difference between the contacts rather than by applied voltage as here) was carried out in Ref. [S11].

### B. Noise on the $\nu = 2/3$ edge

We consider the $\nu = 2/3$ edge hosting two counter-propagating bare modes, the downstream $\nu_+ = 1$ mode and the upstream $\nu_- = 1/3$ mode. The expectation values of the local noise $S_{\text{loc}}$ and the electron tunneling conductance $g_{\text{loc}}$ can be computed within the chiral Luttinger liquid model [S12, S13]. To leading order in the local tunneling strength $\Gamma_0$, these quantities are given as [S10, S14]

$$S_{\text{loc}}(x) = 4 \int_{-\infty}^{\infty} d\tau \cos\left[\frac{e\Delta V(x)\tau}{\hbar}\right] \langle \mathcal{T}(\tau,0)\mathcal{T}^\dagger(0,0) \rangle, \tag{S17}$$

$$g_{\text{loc}}(x) = 2i \int_{-\infty}^{\infty} d\tau \tau \langle \mathcal{T}(\tau,0)\mathcal{T}^\dagger(0,0) \rangle. \tag{S18}$$

Here, the electron tunneling operator

$$\mathcal{T}(\tau,0) = \frac{\Gamma_0}{2\pi b} \exp\left[i\sqrt{\Delta-1}\phi_+(\tau,0) + i\sqrt{\Delta+1}\phi_-(\tau,0)\right], \tag{S19}$$

is written in terms of the two bosonic eigenmodes $\phi_\pm$ in the interacting region [S10], and $b$ is a short distance cutoff. The parameter $\Delta$ quantifies the strength of interactions, with $\Delta = 2$ for vanishing interactions, and $\Delta = 1$ for strong interactions [S15, S16]. Assuming that each of the eigenmodes is in local equilibrium, with respective temperatures $T_+$ and $T_-$, the correlation functions are evaluated as

$$\langle \mathcal{T}(\tau,0)\mathcal{T}^\dagger(0,0) \rangle = \frac{|\Gamma_0|^2}{(2\pi b)^2} G_+(\tau,0)^{2d_+} G_-(\tau,0)^{2d_-}, \tag{S20}$$

with the finite temperature Green's functions

$$G_\pm(\tau,0) = \frac{\pi b T_\pm / v_\pm}{\sin\left[\frac{\pi T_\pm}{v_{L,R}}(b - i\tau v_\pm)\right]}, \tag{S21}$$

and the exponents

$$2d_+ = \Delta - 1, \tag{S22}$$
$$2d_- = \Delta + 1. \tag{S23}$$

The Fourier transforms of the factors $G_\pm(\tau,0)^{2d_\pm}$ read

$$P_\pm(\omega, T_\pm) \equiv \int_{-\infty}^{\infty} d\tau e^{i\omega\tau} G_\pm(\tau,0)^{2d_\pm}$$
$$= \left(\frac{2\pi b T_\pm}{v_\pm}\right)^{2d_\pm - 1} \left(\frac{b}{v_\pm}\right) e^{\omega/(2T_\pm)} \frac{|\Gamma(d_\pm + i\frac{\omega}{2\pi T_\pm})|^2}{\Gamma(2d_\pm)}, \tag{S24}$$

with $\Gamma(z)$ being the gamma-function. We also give the zero temperature limits of Eqs. (S21) and (S24), in which

$$G_\pm(\tau,0) = \frac{b}{(b - i\tau v_\pm)}, \tag{S25}$$

$$P_\pm(\omega, 0) = \frac{2\pi (b/v_\pm)^{2d_\pm} \omega^{2d_\pm - 1} \Theta(\omega)}{\Gamma(2d_\pm)}, \tag{S26}$$

with $\Theta(\omega)$ being the step function. With the Fourier transforms, the noise (S17) and the tunneling conductance (S18) can be expressed as

$$S_{\text{loc}}(x) = \frac{4|\Gamma_0|^2}{2(2\pi b)^2} \int_{-\infty}^{\infty} \frac{d\omega}{2\pi} P_+(-\omega, T_+) \left[ P_-\left(\omega + \frac{e\Delta V}{\hbar}, T_-\right) + P_-\left(\omega - \frac{e\Delta V}{\hbar}, T_-\right) \right], \tag{S27}$$

$$g_{\text{loc}}(x) = \frac{2|\Gamma_0|^2}{(2\pi b)^2} \frac{\partial}{\partial \omega'} \left( \int_{-\infty}^{\infty} \frac{d\omega}{2\pi} P_+(\omega' - \omega, T_+) P_-(\omega, T_-) \right) \bigg|_{\omega'=0}. \tag{S28}$$

For notational ease, we have here suppressed the $x$-dependence on $\Delta V$ and $T_\pm$. Combining Eqs. (S27) and (S28), we express the noise kernel (S16) as

$$\Lambda(x) = \frac{1}{2} \times \frac{\int_{-\infty}^{\infty} d\omega P_+(-\omega, T_+) \left[ P_-(\omega + \frac{e\Delta V}{\hbar}, T_-) + P_-(\omega - \frac{e\Delta V}{\hbar}, T_-) \right]}{\frac{\partial}{\partial \omega'} \left( \int_{-\infty}^{\infty} d\omega P_+(\omega' - \omega, T_+) P_-(\omega, T_-) \right) \bigg|_{\omega'=0}}. \tag{S29}$$

Let us briefly discuss the content of Eq. (S29). First, the noise kernel depends on the interaction strength (parameter $\Delta$) through the exponents (S22)-(S23) entering the Fourier transforms $P_\pm$ in Eq. (S24). The noise kernel has also generally a spatial dependence via the local temperatures $T_\pm(x)$ and the local voltage difference $\Delta V(x)$. We turn now to the computation of these two quantities.

*1. Voltage drop and dissipated power along the edge*

First, we compute voltage profiles along the $\nu = 2/3$ edge to extract the voltage drop $\Delta V(x)$. Charge transport along the edge segment is described by the transport equation [S10, S17]

$$\partial_x \begin{pmatrix} I_1(x) \\ I_2(x) \end{pmatrix} = \frac{1}{l_{\text{eq}}^C} \begin{pmatrix} -1 & 3 \\ -1 & 3 \end{pmatrix} \begin{pmatrix} I_1(x) \\ I_2(x) \end{pmatrix}, \tag{S30}$$

expressed in terms of the local bare-mode charge currents $I_{1,2}(x)$ and the charge equilibration length $l_{\text{eq}}^C$. This equation should be supplemented by boundary conditions at the contacts. The natural physical picture is that the metallic contacts screen the interaction, and one has bare modes coupled to the contact [S16]. The corresponding voltages are thus applied to the bare modes emanating form the contact, yielding the boundary conditions $I_1(0) = e^2/h \times V_0$ and $I_2(L) = 0$ (see Ref. S17 for a recent detailed discussion regarding such boundary conditions). Solving Eq. (S30) and using the relations $I_1(x) = e^2 V_1(x)/h$ and $I_2(x) = e^2 V_2(x)/3h$, we find the local voltage drop

$$\Delta V(x) \equiv [V_1(x) - V_2(x)] = V_0 \left[ \frac{2}{3} \frac{e^{2x/l_{\text{eq}}^C}}{e^{2L/l_{\text{eq}}^C} - 1/3} \right]. \tag{S31}$$

Equation (S31) reveals that in the limit $L \gg l_{\text{eq}}^C$, the voltage drop occurs only in a region of size $\sim l_{\text{eq}}^C$ close to the right contact (see Fig. S10), $L - x \lesssim l_{\text{eq}}^C$. We call this the hot spot since this is where the Joule heating takes place. In the rest of the segment, $\Delta V(x)$ is exponentially suppressed. In particular, at the noise spot ($x \lesssim l_{\text{eq}}^C$), the voltage drop is exponentially small $[\sim \exp(-2L/l_{\text{eq}}^C)]$ and can be safely approximated by zero.

We also compute the power $P$ dissipated in the hot spot. Under the assumption of strong charge equilibration, electrical energy conservation in the edge modes leads to the formula [S8]

$$P = \frac{e^2 V_0^2}{2h} \times \frac{(\nu_+ - \nu_-)\nu_-}{\nu_+}. \tag{S32}$$

Applied to $\nu = 2/3$, where $\nu_+ = 1$ and $\nu_- = 1/3$, Eq. (S32) gives

$$P_{2/3} = \frac{e^2 V_0^2}{9h}. \tag{S33}$$

This power becomes distributed among the edge channels and then propagates towards both contacts limiting the edge segments. (We assume no dissipation to the environment.) The resulting profile of effective temperatures $T_\pm(x)$ depend on the degree of thermal equilibration within the edge. Below we compute $T_\pm$ in the regime of negligibly weak thermal equilibration.

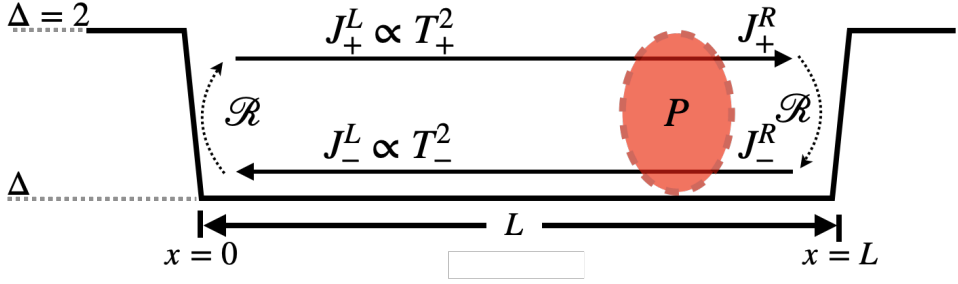

FIG. S11: Determination of steady-state temperatures $T_\pm$ on the edge (to the left of the hot spot). The dissipated power (Joule heat) $P$ in the hot spot is distributed among left- and right-propagating eigenmodes. At interfaces with contact regions (near the points $x = 0$ and $x = L$), the heat is partially reflected (reflection coefficient $\mathcal{R}$) due to different strength of interaction in the central region (strong interaction, parameter $\Delta$) and in contact regions (no interaction, $\Delta = 2$). This reduces the energy escape from the edge into the contacts, thus enhancing the the temperatures $T_\pm$. The steady state is determined by the system of equations (S34), which yield the temperatures (S36) and (S37).

### 2. Effective temperatures

We compute now the effective temperatures $T_\pm$. Assuming vanishing thermal equilibration, thermal transport on the edge can be modelled as in Fig. S11. The dissipated power $P_{2/3}$ [see Eq. (S33)] acts as a local heat source for the edge heat currents. To find the temperatures of the edge modes, we consider four heat currents $J_\pm^{L,R} = 0.5 \times \kappa_0 (T_\pm^{L,R})^2$ [with $\kappa_0 = \pi^2 k_B^2/(3h)$], which are right $(+)$ and left $(-)$ moving currents to the left $(L)$ and to the right $(R)$ of the heat source as depicted in Fig. S11. Power conservation then leads to the following set of equations:

$$J_+^L = \mathcal{R} J_-^L, \tag{S34a}$$
$$J_-^R = \mathcal{R} J_+^R, \tag{S34b}$$
$$J_+^R - J_+^L = P_+, \tag{S34c}$$
$$J_-^L - J_-^R = P_-. \tag{S34d}$$

Here, the first two equations describe the reflection of the heat currents due to the change in interaction strength at the contacts. Specifically, away from the contacts, the interaction is strong and characterized by the parameter $\Delta$, while in the contact region we have no interaction, which corresponds to $\Delta = 2$. This reflection is parametrized by the reflection coefficient $\mathcal{R}$, given in terms of the interaction parameter $\Delta$ via the relations [S16]

$$\mathcal{R} = \left(\frac{1 - \sqrt{1-c^2}}{c}\right)^2, \quad \Delta = \frac{2 - \sqrt{3}c}{\sqrt{1-c^2}}. \tag{S35}$$

Here $c$ is proportional to the strength of interaction and satisfies (assuming a repulsive interaction) $0 \le c < 1$. For the strong-interaction fixed point $\Delta = 1$, we have $c = \sqrt{3}/2$, and $\mathcal{R} = 1/3$. (For vanishing interactions, $c = 0$, one has $\Delta = 2$ and $\mathcal{R} = 0$: there is no reflections when the interaction vanishes both in the contacts and in the central region.)

Since the $\nu = 2/3$ edge hosts one bosonic mode in each direction, we assume that the dissipated power is distributed equally for right and left moving currents, i.e., $P_+ = P_- = 0.5 \times P_{2/3}$ with $P_{2/3}$ given in Eq. (S33). Then, by solving Eq. (S34), we find the effective eigenmode temperatures to the left of the heat source:

$$k_B T_+ \equiv k_B T_+^L = \frac{eV_0}{\pi\sqrt{3}} \left(\frac{\mathcal{R}}{1-\mathcal{R}}\right)^{1/2}, \tag{S36}$$

$$k_B T_- \equiv k_B T_-^L = \frac{eV_0}{\pi\sqrt{3}} \left(\frac{1}{1-\mathcal{R}}\right)^{1/2}. \tag{S37}$$

At the strongly interacting fixed point ($\Delta = 1$), we have $\mathcal{R} = 1/3$, which yields

$$k_B T_+ = 0.13 eV_0, \qquad k_B T_- = 0.23 eV_0. \tag{S38}$$

Equations (S36)-(S37) indicate that reflections at the contact interfaces tend to increase the overall temperature on the edge. (For example, in the limit of absent inter-mode interaction, when $\mathcal{R} = 0$, we would have $k_B T_+ = 0$ and $k_B T_- = 0.18 eV_0$.) This is quite transparent physically, since the reflections reduce the escape of the heat from the edge into the contacts. Note that the temperatures $T_+$, $T_-$ given by Eqs. (S36), (S37), (S38) are independent on the coordinate $x$ (within the segment of the edge from the left contact to the hot spot), which is a manifestation of the absence of thermal equilibration in the edge. These formulas thus give, in particular, the effective temperatures at the noise spot, which are needed to calculate the noise as given by Eq. (S15). We proceed now with this calculation.

### 3. Results

We now have all the ingredients that are needed to compute the noise (S15) for the $\nu = 2/3$ edge. For the voltage drop, Eq. (S31) yields yields an exponentially small $\Delta V$ at the noise spot, and we can thus safely set it to zero. Further, we use Eqs. (S36)-(S37) for the temperatures of both eigenmodes. Since $\Lambda(x)$ in Eq. (S29) is independent of $x$ in the considered regime of vanishing thermal equilibration, we can trivially perform the $x$-integration in Eq. (S15). We then arrive at

$$S_{2/3} = \frac{2e^2}{9h} \Lambda(V_0, \Delta), \tag{S39}$$

where we have emphasized the dependence of $\Lambda$ on the interaction parameter $\Delta$ and on the bias voltage $V_0$. The interaction $\Delta$ enters via the exponents (S22)–(S23) and via the reflection coefficent (S35). The voltage bias $V_0$ enters via the local temperatures (S36)–(S37).

We express now the noise $S_{2/3}$ in terms of the bias current

$$I_0 = \frac{2e^2}{3h} V_0. \tag{S40}$$

Further, we take the value $\Delta = 1$ corresponding to the strong-interaction fixed point. Experimental observation of the hierarchy of equilibration lengths, $l_{eq}^C \ll l_{eq}^H$, indicates that $\Delta$ is close to unity as discussed above. Equation (S39), in combination with Eq. (S29) for $\Lambda$, then yields

$$S_{2/3} = 0.146 \, eI_0. \tag{S41}$$

We have checked that small deviations of $\Delta$ from the value $\Delta = 1$ lead only to very small variations of the numerical prefactor in Eq. (S41).

Equation (S41) represents our key result for the bias-induced noise on a $\nu = 2/3$ edge. Let us emphasize that the noise (S41) is independent on the length $L$, which is hallmark of the thermally non-equilibrated regime, $L \ll l_{eq}^H$. In the opposite limit of strong thermal equilibration ($L \gg l_{eq}^H$), the noise is suppressed and decays as $(L/l_{eq}^H)^{-1/2}$ [S6, S7]. The experimentally observed $L$-independence of the noise (Fig. 2d of the main manuscript) thus provides a clear evidence of the thermally non-equilibrated regime. The theoretical formula (S41) is compared with our experimental results in Fig. 2d of the main manuscript and in Fig. S8b of this Supplementary Material. The excellent agreement between the theoretical and experimental magnitude of the noise provides further strong support to our interpretation of the experiment.

### C. Noise at filling 3/5

The edge structure at $\nu = 3/5$ is more complicated than at $\nu = 2/3$, since the edge of a hole-conjugated $\nu = 3/5$ state hosts three bare edge modes: $\phi_j$ with $j = 1, 2, 3$ and $\nu_1 = 1$, $\nu_2 = 1/3$, and $\nu_3 = 1/15$. Whereas $\phi_1$ is propagating downstream, $\phi_2$ and $\phi_3$ are upstream modes [S18, S19].

To capture the essential physics on this edge, we apply three approximations. First, we note that charge tunneling between the two co-propagating upstream modes can not partition charges between the two contacts. Such processes therefore do not influence the noise and can be ignored. Second, we assume an efficient equilibration between the two upstream modes, so that the edge essentially consists of two effective counter-propagating, hydrodynamic modes with $\nu_+ = \nu_1 = 1$ and $\nu_- = \nu_2 + \nu_3 = 1/3 + 1/15 = 2/5$ (see Fig. S10). Third, in order to simplify the technical analysis, we assume that electron tunneling between the adjacent $\phi_1$ and $\phi_2$ modes is much stronger than that between $\phi_1$ and $\phi_3$ which are further apart. The renormalization of these two tunneling processes follow the same renormalization group equations [S19], so if the difference in tunneling amplitudes is large at some energy, it will remain large during

renormalization to lower energies. In principle, any of this approximations can be relaxed. While this would make calculations substantially more cumbersome, one would end up with nearly the same result, up to a small variation of the numerical prefactor.

With these simplifications, the microscopic description of the noise kernel $\Lambda(x)$ [see Eq. (S16)] becomes exactly the same as the one for the $\nu = 2/3$ edge, i.e., it is given by Eq. (S29). What remains to be computed for the $\nu = 3/5$ edge is then the voltage drop $\Delta V(x)$ and the effective temperatures $T_\pm$ of downstream and upstream eigenmodes in the interacting region. (Here $T_-$ is the temperature of both upstream eigenmodes since, in order to simplify the analysis, they are assumed above to equilibrate efficiently.)

### 1. Voltage drop and dissipated power along the edge

The analysis of Sec. S9 B 1 is straightforwardly extended to the $\nu = 3/5$ edge. In full analogy with to the $\nu = 2/3$ edge, the voltage drop at the $\nu = 3/5$ edge occurs only close to the right contact and is exponentially suppressed away from it [S7]. In particular, the voltage drop $\Delta V(x)$ is exponentially suppressed in $L/l_{eq}^C$ at the noise spot (near the left contact) and can be approximated by zero.

To compute the dissipated power, we apply Eq. (S32) with $\nu_+ = 1$ and $\nu_- = 2/5$. The result is

$$P_{3/5} = \frac{3e^2 V_0^2}{25h}. \tag{S42}$$

We note that dissipated powers $P_{2/3}$ [see Eq. (S33)] and $P_{3/5}$ are very close in magnitude.

### 2. Effective temperatures

We consider four heat currents $J_\pm^{L,R} = 0.5 \times n_\pm \times \kappa_0 (T_\pm^{L,R})^2$, denoting right $(+)$ and left $(-)$ moving currents to the left $(L)$ and to the right $(R)$ of the heat source (see Fig. S11). At variance with the $\nu = 2/3$ edge, we need here to take into account the difference in the number of downstream and upstream modes. This is done by including the numbers $n_+ = 1$ and $n_- = 2$ in the heat currents above. Power conservation on the $\nu = 3/5$ edge leads to the following system of equations that represents an extension of the system (S34):

$$J_+^L = \frac{1}{2}\mathcal{R} J_-^L, \tag{S43a}$$

$$J_-^R = \mathcal{R} J_+^R, \tag{S43b}$$

$$J_+^R - J_+^L = P_+, \tag{S43c}$$

$$J_-^L - J_-^R = P_-. \tag{S43d}$$

Note the factor of $1/2$ in Eq. (S43a) as compared to Eq. (S34a). This is a consequence of the difference in the number of downstream and upstream modes.

We assume that the dissipated power $P_{3/5}$ [given by Eq. (S42)] is distributed proportionally to the number of downstream and upstream modes ($n_+$ and $n_-$), so that $P_+ = (1/3) P_{3/5}$ and $P_- = (2/3) P_{3/5}$. Further, the reflection coefficient $\mathcal{R}$ depends on the interactions between the modes of the $3/5$ edge. In analogy with the above analysis of the $2/3$ edge, we assume that the system is close to the strong-interaction fixed point (counterpart of $\Delta = 1$ point of the $2/3$ edge). As discussed above, this assumption is consistent with experimental observation of nearly vanishing thermal equilibration. At the strongly interacting fixed point, the reflection coefficient for the $3/5$ edge reads [S11]

$$\mathcal{R} = 1 - \nu = 2/5. \tag{S44}$$

The solution to (S43) for the temperatures to the left of the hot spot (in particular, at the the noise spot) then becomes

$$k_B T_+ \equiv k_B T_+^L = \sqrt{\frac{2}{23}} \times \frac{6eV_0}{5\pi} \approx 0.11 eV_0, \tag{S45}$$

$$k_B T_- \equiv k_B T_-^L = \sqrt{\frac{1}{115}} \times \frac{6eV_0}{\pi} \approx 0.18 eV_0. \tag{S46}$$

Both temperatures are somewhat below their values for the $2/3$ edge, Eq. (S38), but the differences are not too big. For comparison, if one assumes no interaction on the edge, and thus $\mathcal{R} = 0$, one gets the temperatures $k_B T_+ = 0$ and $k_B T_- \approx 0.16 eV_0$.

### 3. Digression: Temperatures in the case of a generic edge

As a short digression, we compute the effective temperatures $T_\pm$ for the case of generic $n_\pm$ by extending the analysis of Ref. S9 B 2 and S9 C 2. We consider four heat currents $J_\pm^{L,R} = 0.5 \times n_\pm \times \kappa_0 (T_\pm^{L,R})^2$ [with $\kappa_0 = \pi^2 k_B^2/(3h)$] denoting right (+) and left (−) moving currents to the left ($L$) and to the right ($R$) of the heat source. We assume that all downstream (upstream) modes have the same temperature $T_+$ (respectively, $T_-$). Power conservation then leads to the following set of equations:

$$J_+^L = \frac{\mathcal{R}}{n_-} J_-^L, \tag{S47a}$$

$$J_-^R = \frac{\mathcal{R}}{n_+} J_+^R, \tag{S47b}$$

$$J_+^R - J_+^L = P_+ = P \times \frac{n_+}{n_+ + n_-}, \tag{S47c}$$

$$J_-^L - J_-^R = P_- = P \times \frac{n_-}{n_+ + n_-}. \tag{S47d}$$

Solving for $J_\pm^L$, we obtain

$$J_+^L = P \times \frac{\mathcal{R} n_+ (n_- + \mathcal{R})}{(n_+ + n_-)(n_+ n_- - \mathcal{R}^2)}, \tag{S48}$$

$$J_-^L = P \times \frac{n_+ n_- (n_- + \mathcal{R})}{(n_+ + n_-)(n_+ n_- - \mathcal{R}^2)}. \tag{S49}$$

Converting these currents to temperatures, we find

$$k_B T_+^L = \left( \frac{6hP}{\pi^2} \times \frac{\mathcal{R}(n_- + \mathcal{R})}{(n_+ + n_-)(n_+ n_- - \mathcal{R}^2)} \right)^{1/2}, \tag{S50}$$

$$k_B T_-^L = \left( \frac{6hP}{\pi^2} \times \frac{n_+(n_- + \mathcal{R})}{(n_+ + n_-)(n_+ n_- - \mathcal{R}^2)} \right)^{1/2}. \tag{S51}$$

For $P = P_{2/3} = e^2 V_0^2/9h$, $n_+ = 1$, $n_- = 1$, and $\mathcal{R} = 1/3$, we recover $k_B T_+^L \approx 0.13 eV_0$ and $k_B T_-^L \approx 0.23 eV_0$, Eq. (S38). For $P = P_{3/5} = 3e^2 V_0^2/25h$, $n_+ = 1$, $n_- = 2$ and $\mathcal{R} = 2/5$, we recover $k_B T_+^L \approx 0.11 eV_0$ and $k_B T_-^L \approx 0.18 eV_0$, Eqs. (S45) and (S46).

### 4. Results for the noise on the $\nu = 3/5$ edge

After a short digression in Sec. S9 C 3, we finalize the calculation of the noise for $\nu = 3/5$. Performing the integration over $x$ (which is trivial due to constant temperatures in the noise spot) in Eq. (S15), we obtain for the noise on the $\nu = 3/5$ edge

$$S_{3/5} = \frac{6e^2}{25h} \Lambda(V_0, \Delta), \tag{S52}$$

which is a counterpart of the $\nu = 2/3$ formula (S39). For the strong-interaction fixed point ($\Delta = 1$), the the temperatures are given by Eqs. (S45) and (S46). Using them in Eq. (S29), we determine $\Lambda$, which yields, by virtue of Eq. (S52), the noise $S_{3/5}$. Expressing it in terms of the bias current

$$I_0 = \frac{3e^2}{5h} V_0, \tag{S53}$$

we finally obtain

$$S_{3/5} = 0.138 \, eI_0. \tag{S54}$$

Note that the noise is very close to that on the 2/3 edge, Eq. (S41). The theoretical result (S54) agrees very well with our experimental data for the noise on the 3/5 edge, see Fig. S8c.

It is interesting to compare the result (S54) obtained for the thermally non-equilibrated 3/5 edge ($L \gg l_{\text{eq}}^H$) to the noise $S_{3/5,\text{eq}}$ found on the same edge in the opposite limit of strong thermal equilibration ($L \ll l_{\text{eq}}^H$). The latter calculation was carried out in Ref. S7, where it was found that

$$S_{3/5,\text{eq}} = 0.116\, eI_0. \tag{S55}$$

We see that the magnitudes of noise in the two limits differ by approximately 20%. The difference is much less dramatic than in the case of the $\nu = 2/3$ noise (which is parametrically suppressed in the thermally equilibrated regime). This is because, in the regime of strong thermal equilibration, the upstream heat transport remains ballistic on the 3/5 edge but becomes diffusive on the 2/3 edge.

---



\end{document}